\documentclass{aa}  

\usepackage{graphicx}
\usepackage{txfonts}
\usepackage{multirow}
\usepackage{hyperref}
\usepackage{mathrsfs}
\usepackage{xcolor}

\begin{document}

   \title{The CARMENES search for exoplanets around M dwarfs}

   \subtitle{The \ion{He}{i} infrared triplet lines in PHOENIX models of M2--3\,V stars}

   \author{D.~Hintz\inst{1}
          \and  B.~Fuhrmeister\inst{1}
          \and  S.~Czesla\inst{1}
          \and  J.~H.~M.~M.~Schmitt\inst{1}
          \and  A.~Schweitzer\inst{1}
          \and  E.~Nagel\inst{2}
          \and  E.~N.~Johnson\inst{3}
          \and  J.~A.~Caballero\inst{4}
          \and  M.~Zechmeister\inst{3}
          \and  S.~V.~Jeffers\inst{3}
          \and  A.~Reiners\inst{3}
          \and  I.~Ribas\inst{5,6}
          \and  P.~J.~Amado\inst{7}
          \and  A.~Quirrenbach\inst{8}
          \and  G.~Anglada-Escud\'e\inst{7,9}
          \and  F.~F.~Bauer\inst{7}
          \and  V.~J.~S.~B\'ejar\inst{10,11}
          \and  M.~Cort\'es-Contreras\inst{4}
          \and  S.~Dreizler\inst{3}
          \and  D.~Galad\'{\i}-Enr\'{\i}quez\inst{12}
          \and  E.~W.~Guenther\inst{2,10}
          \and  P.~H.~Hauschildt\inst{1}
          \and  A.~Kaminski\inst{8}
          \and  M.~K\"urster\inst{13}
          \and  M.~Lafarga\inst{5,6}
          \and  M.~L\'opez~del~Fresno\inst{4}
          \and  D.~Montes\inst{14}
          \and  J.~C.~Morales\inst{5,6}
          }

   \institute{Hamburger Sternwarte, University of Hamburg, 
              Gojenbergsweg 112, D-21029 Hamburg, Germany \\
              \email{dominik.hintz@hs.uni-hamburg.de}
         \and
              Th\"uringer Landessternwarte Tautenburg, Sternwarte 5, D-07778 Tautenburg, Germany 
         \and
              Institut für Astrophysik, 
              Friedrich-Hund-Platz 1, D-37077 Göttingen, Germany
         \and 
              Centro de Astrobiología (CSIC-INTA), ESAC, 
              Camino Bajo del Castillo s/n, E-28692 Villanueva de la Cañada, Madrid, Spain
         \and
              Institut de Ci\`encies de l'Espai (ICE, CSIC), Campus UAB, c/ de Can Magrans s/n, E-08193 Bellaterra, Barcelona, Spain
         \and
              Institut d'Estudis Espacials de Catalunya (IEEC), E-08034 Barcelona, Spain
         \and 
              Instituto de Astrof\'isica de Andaluc\'ia (CSIC), Glorieta de la Astronom\'ia s/n, E-18008 Granada, Spain 
         \and 
              Landessternwarte, Zentrum f\"ur Astronomie der Universit\"at Heidelberg, K\"onigstuhl 12, D-69117 Heidelberg, Germany 
         \and
              School of Physics and Astronomy, Queen Mary, University of London, 327 Mile End Road, London, E1 4NS, UK
         \and
              Instituto de Astrof\'{\i}sica de Canarias, c/ V\'{\i}a L\'actea s/n, E-38205 La Laguna, Tenerife, Spain
         \and
              Departamento de Astrof\'{\i}sica, Universidad de La Laguna, E-38206 Tenerife, Spain 
         \and
              Centro Astron\'omico Hispano-Alem\'an (MPG-CSIC), Observatorio Astron\'omico de Calar Alto, Sierra de los Filabres, E-04550 G\'ergal, Almer\'{\i}a, Spain 
         \and
              Max-Planck-Institut f\"ur Astronomie, K\"onigstuhl 17, D-69117 Heidelberg, Germany 
         \and
              Departamento de F\'{i}sica de la Tierra y Astrof\'{i}sica and UPARCOS-UCM (Unidad de F\'{i}sica de Part\'{i}culas y del Cosmos de la UCM), 
              Facultad de Ciencias F\'{i}sicas, Universidad Complutense de Madrid, E-28040, Madrid, Spain
             }

   \date{Received 28 January 2020 / accepted dd mm 2020}

  \abstract
  {
   The \ion{He}{i} infrared (IR) line at a vacuum wavelength of 10\,833$\,\AA$ 
   is a diagnostic for the 
   investigation of atmospheres of stars and planets orbiting them. 
   For the first time, we study the behavior of the \ion{He}{i}~IR line in a set of chromospheric models for M-dwarf stars, whose much denser
   chromospheres may favor collisions for the level population over 
   photoionization and recombination, which are believed to be dominant in solar-type stars. 
   For this purpose, we use published PHOENIX models for stars of spectral types M2\,V and M3\,V 
   and also compute new series of models with different levels of activity
   following an ansatz developed for the case of the Sun. 
   We perform a detailed analysis of the behavior of the \ion{He}{i}~IR line within these models. 
   We evaluate the line in relation to other chromospheric lines and also the influence 
   of the extreme ultraviolet (EUV) radiation field. 
   The analysis of the \ion{He}{i}~IR line strengths as a function of  
   the respective EUV radiation field strengths 
   suggests that the mechanism of photoionization and recombination is 
   necessary to form the line for inactive models, while collisions start to
   play a role in our most active models. 
   Moreover, the published model set, which is optimized in the ranges of the 
   \ion{Na}{i}~D$_2$, H$\alpha$, and the bluest \ion{Ca}{ii} IR 
   triplet line, gives an adequate prediction of the \ion{He}{i}~IR line for most stars 
   of the stellar sample. Because especially the most inactive stars with weak \ion{He}{i}~IR lines
   are fit worst by our models, it seems that our assumption of a 100\% 
   filling factor of a single inactive component no longer holds for these stars.  
   }


   \titlerunning{The \ion{He}{i} IR line in PHOENIX models of M2--3\,V stars}
   \maketitle

\section{Introduction}
   Spectral lines arising in a stellar chromosphere provide essential information 
   on stellar activity. In the optical wavelength range, 
   widely used chromospheric indicator lines include the 
   \ion{Ca}{ii}~H and K lines (3934.77$\,\AA$, 3969.59$\,\AA$),
   the \ion{He}{i}~D$_3$ line (5877.25$\,\AA$), 
   the \ion{Na}{i}~D doublet (5897.56$\,\AA$, 5891.58$\,\AA$), 
   and the H$\alpha$ (6564.62$\,\AA$) line. 
   In the infrared, the \ion{Ca}{ii} infrared triplet (IRT) lines 
   (8500.35$\,\AA$, 8544.44$\,\AA$, 8664.52$\,\AA$) 
   and the \ion{He}{i} 
   IRT lines (hereafter \ion{He}{i}~IR line
   for short) at 10\,833$\,\AA$ 
   are widely used chromospheric diagnostics.\footnote{We use 
   vacuum wavelengths throughout the paper. However, 
   the \ion{He}{i}~IR line is well known for its air wavelength at 10\,830$\,\AA$.} 
   Specifically, the \ion{He}{i}~IR line has recently become even more important 
   in the light of investigations of exoplanet atmospheres 
   around late-type stars 
   \citep[e.g., ][]{Spake2018Natur.557...68S, Nortmann2018Sci...362.1388N, Salz2018A&A...620A..97S}. 
   For this reason, it is imperative to improve our understanding of 
   the \ion{He}{i}~IR line behavior of stars, especially for the very late-type stars. 
   In this paper, we investigate the behavior of the \ion{He}{i}~IR line 
   in early M-dwarf stars. 
   
   The \ion{He}{i}~IR line is a triplet line of orthohelium, whose 
   three components are centered at 10\,832.057, 10\,833.217, and 
   10\,833.306$\,\AA$. 
   The term scheme of the neutral helium atom splits into
   the ortho- and parahelium branch, depending on whether
   the electrons have aligned or opposite spins. 
   Singlet lines only occur in parahelium, while 
   orthohelium is characterized by triplet lines. 
   
   The lower level of the \ion{He}{i}~IR line is the metastable $2\,^3S$ level of orthohelium, 
   which is located about 20\,eV above the ground state of (para-)helium. 
   The \ion{He}{i}~IR line appears in absorption when 
   the occupation number of the metastable level is higher than that of the overlying 
   $2\,^3P$ triplet levels, and vice versa for an emission line. 
   A decisive question is which process is responsible for populating the metastable ground state. 
   Previous studies modeling the \ion{He}{i}~IR line 
   \citep[e.g.,][]{Andretta1995ApJ...439..405A, Andretta1997ApJ...489..375A} investigated  
   two possible line formation mechanisms in detail: (i) the photoionization and recombination (PR)
   mechanism, and (ii) collisional excitation. 
   
   In the PR formation scenario,  
   the continuum shortward of $504\,\AA$ is important for photoionization 
   of \ion{He}{i} in the ground state. 
   The photons from this continuum,  
   produced in the transition region and corona,
   photoionize the \ion{He}{i} atoms and produce free electrons. 
   These can be captured by the helium ion into a highly excited energy level, from which they finally 
   cascade down to the metastable $2\,^3S$ level. The excited \ion{He}{i} atom then interacts with photospheric photons. 
   The PR mechanism is thought to dominate the population of the metastable level 
   at temperatures below $10\,000\,$K in the upper chromosphere. 
   In the collisional line formation scenario, a helium atom in the ground state is excited by an electron 
   collision to the $2\,^3S$ level. This process requires a sufficiently dense high-energy electron population, which is typically only encountered
   in the hot upper chromosphere and lower transition region at temperatures above $20\,000$\,K. 
   
   \citet{Vaughan1968ApJ...152..123V} were the first to 
   study the \ion{He}{i}~IR line in stars of spectral type between F and M. In their study, 
   \ion{He}{i}~IR absorption was detected for most of the G and early K-type stars in their sample. 
   Further observations by \citet{Zirin1975STIN...7614992Z} revealed \ion{He}{i}~IR emission lines 
   for some late M-type stars. 
   \citet{Zirin1975ApJ...199L..63Z} constructed a solar model 
   incorporating the PR mechanism
   and concluded that
   this mechanism is decisive for the formation of the \ion{He}{i}~D$_3$ and \ion{He}{i}~IR lines. 
   Furthermore, they postulated a correlation between the \ion{He}{i}~IR line equivalent widths and the soft X-ray flux 
   for stars that show the line in absorption. This was later observed by  
   \citet{Zarro1986ApJ...304..365Z} for stars in the spectral range between F7 and K3. 
   
   In a study of the formation of the \ion{He}{i}~IR line in the Sun, \citet{Avrett1994IAUS..154...35A} generated different 
   solar models based on the VAL~C model by \citet{Vernazza1981} for the average quiet Sun. 
   In varying the atmospheric height of the transition region onset as well as changing the incident radiation from the transition region, 
   \citet{Avrett1994IAUS..154...35A} found that enhancing the radiation leads to an increase in absorption of 
   the solar \ion{He}{i}~IR line.
   In contrast, a solar plage model, characterized by a radiation level thrice that 
   of the quiet Sun and a transition region moved inward, 
   showed a slightly weaker \ion{He}{i}~IR line than the average quiet model. 
   The authors attributed the weakening of the line strength in the plage model to its high density and to the relatively lower 
   geometric extension of the chromosphere in this model. 
   \citet{Andretta1995ApJ...439..405A} found a clear correlation between the X-ray emission and the 
   neutral helium lines in chromospheric models of F, G, and early K stars  by investigating the behavior 
   of the \ion{He}{i}~D$_3$ and \ion{He}{i}~IR line with the VAL~C model shifted in density. 
   Intensifying the ionizing radiation field leads to stronger \ion{He}{i}~IR line absorption 
   until a limit of  $\sim$400\,m$\AA$ in the equivalent width is reached. Then 
   the behavior reverses and the line tends to go into emission. Studying the \ion{He}{i}~IR line of 
   stars of spectral types G, K, and M, 
   \citet{Sanz-Forcada2008A&A...488..715S} concluded that the collisional excitation mechanism may 
   contribute to the line formation for active dwarfs because high densities of neutral helium in the 
   upper chromosphere can be reached. This conclusion agrees with the results of \citet{Andretta1995ApJ...439..405A}. 
   
   \citet{Andretta1997ApJ...489..375A} theoretically studied the effects of these 
   two \ion{He}{i}~IR line formation mechanisms on the solar spectrum. 
   From computing a series of solar models that varied in density and radiation field, 
   they found that the formation mechanisms influence the \ion{He}{i}~IR line quite differently. 
   While the \ion{He}{i}~IR line is highly sensitive to 
   irradiation from the transition region and corona in the classical VAL~C model for the quiet Sun, 
   collisions become more important when 
   the chromospheric structure is shifted inward toward increasing densities. 
   After an inward shift of ten times the mass load above the quiet-Sun model,  
   the line becomes almost insensitive to the radiation field. Further shifting toward 
   higher densities drives the line into emission.    
   Later, \citet{Leenaarts2016A&A...594A.104L} 
   investigated the spatial structure of the \ion{He}{i}~IR line using a three-dimensional radiation-magnetohydrodynamic simulation. 
   From their model atmosphere, which included the solar chromosphere and corona, they were able to confirm 
   that the PR mechanism is responsible for populating the $2\,^3S$ level in the case of the Sun. 
   Compared to this, the collisional excitation appeared to be negligible. 
   However, they also found that variations of the chromospheric electron density have 
   a crucial effect on variations of the \ion{He}{i}~IR line. 
   
   The number of \ion{He}{i}~IR line studies in M~dwarfs falls far short of that carried out for
   solar-type stars. 
   A recent observational study by \citet{Fuhrmeister2019A&A...632A..24F} used high-resolution infrared spectra obtained with the 
   Calar Alto high-Resolution search for M dwarfs with Exo-earths with 
   Near-infrared and optical Echelle Spectrographs \citep[CARMENES; ][]{Quirrenbach2018SPIE10702E..0WQ} 
   to examine the behavior of the \ion{He}{i}~IR line in M~dwarfs. 
   They found the line to be prominent in early M dwarfs with a tendency to 
   decrease in strength toward later spectral types. 
   Moreover, the line was observed transiently in emission during flaring events.
   \citet{Fuhrmeister2019A&A...632A..24F} was unable to detect a correlation of the line strength
   measured by its pseudo-equivalent width (pEW) and the fractional X-ray luminosity 
   ($L_{\rm X}/L_{\rm bol}$) even for the earliest M dwarfs. 
   It remains unclear whether this noncorrelation is caused by observational deficits
   (e.g., the lack of simultaneous observations) or a stronger effect   of the collisional level population mechanism. \citet{Hintz2019A&A...623A.136H} compared models of optical chromospheric lines
   with observed spectra of a stellar sample of M2--3\,V stars. 
   In the best-fit models, 
   the onset of the transition region 
   can reach column mass densities more than one magnitude higher than the density of the solar VAL~C model 
   even in the low-activity regime, implying much higher electron densities than in the Sun. 
   Even in early M~dwarfs, collisions may therefore play a larger role in the
   \ion{He}{i}~IR line formation. 
   For this reason, chromospheric models dedicated to M~dwarfs need to be computed 
   to theoretically explain the observed behavior.
   
   In this study, we investigate the behavior of the \ion{He}{i}~IR line 
   using a model set of chromospheres based on the model set presented by 
   \citet{Hintz2019A&A...623A.136H}. 
   We introduce the model set as well as the comparison observations, 
   and describe the method of measuring the equivalent widths in Sect.~2. 
   In Sect.~3 we show and discuss our results, and we present our 
   conclusions in Sect.~4.

\section{Models and observations} \label{Models_obs}

\subsection{Observations and stellar sample}
   The observed spectra to which we compared our models were taken with 
   the CARMENES spectrograph \citep{Quirrenbach2018SPIE10702E..0WQ} and retrieved from 
   the CARMENES archive; 
   the CARMENES spectrograph covers the \ion{He}{i}~IR line 
   in its near-infrared channel (NIR), which ranges from $9600$ to $17\,100\,\AA$ with a 
   resolution of $R = 80\,400$. 
   All spectra were reduced by the CARMENES data reduction pipeline \citep{Caballero2016SPIE.9910E..0EC, Zechmeister2018A&A...609A..12Z}. 
   
   For further correction and averaging of the spectra, we followed the same 
   procedures as \citet{Fuhrmeister2019A&A...632A..24F}: 
   The stellar spectra were corrected for telluric contamination 
   using the method of \citet{Nagel2020phd} 
   and for airglow lines by coadding the
   spectra  using the SpEctrum Radial Velocity AnaLyser \citep[SERVAL,][]{Zechmeister2018A&A...609A..12Z}. 
   Furthermore, barycentric and radial velocity shifts were corrected in the spectra. 
   
   We compared the model set to 
   50 M2--3\,V stars, which are 
   a subsample of the stars used by \citet{Fuhrmeister2019A&A...632A..24F}.
   Our sample is defined by the effective temperature range $T_{\rm eff} = 3500 \pm 50\,$K, 
   and has previously been investigated by \citet{Hintz2019A&A...623A.136H}. 
   It comprises four 
   active stars in which the lines of \ion{Na}{i}~D$_2$, H$\alpha$, 
   and the \ion{Ca}{ii}~IRT lines usually appear in emission; 
   for further details about the stellar parameters of the sample, we refer to 
   Table~1 in \citet{Hintz2019A&A...623A.136H}. 
   
\subsection{Measuring the \ion{He}{i} IR line pseudo-equivalent width}
   To characterize the line strengths in the investigated spectra, 
   we calculated the pEW with the same ansatz as \citet{Fuhrmeister2019A&A...632A..24F},
   who fit Voigt profiles to 
   the \ion{He}{i}~IR line profiles. This line profile fitting has a long tradition. It has also been performed,
   for example, by \citet{Takeda2011PASJ...63S.547T} using Gaussian models 
   to estimate the equivalent width of the \ion{He}{i}~IR line in late-type stars. 
   
\begin{figure}
   \centering
   \includegraphics[width=0.4\textwidth]{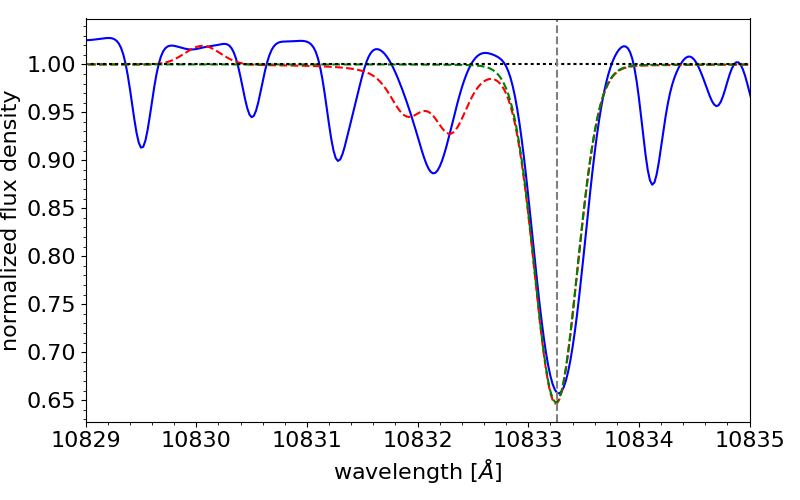}
   \includegraphics[width=0.4\textwidth]{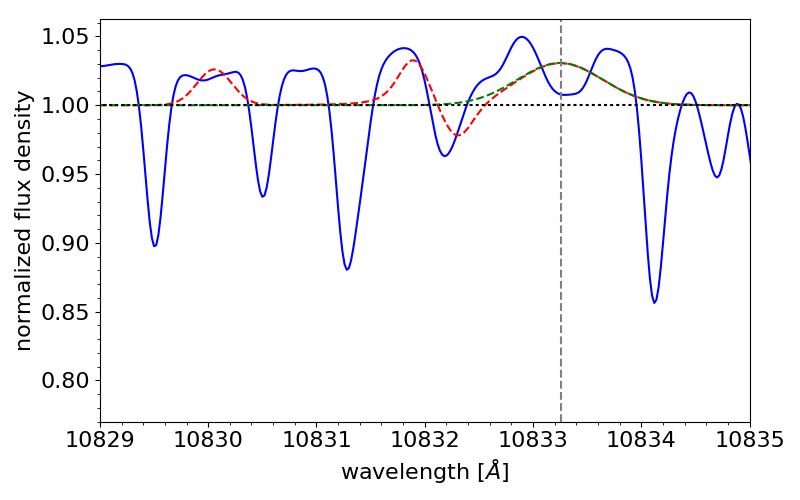}
   \includegraphics[width=0.4\textwidth]{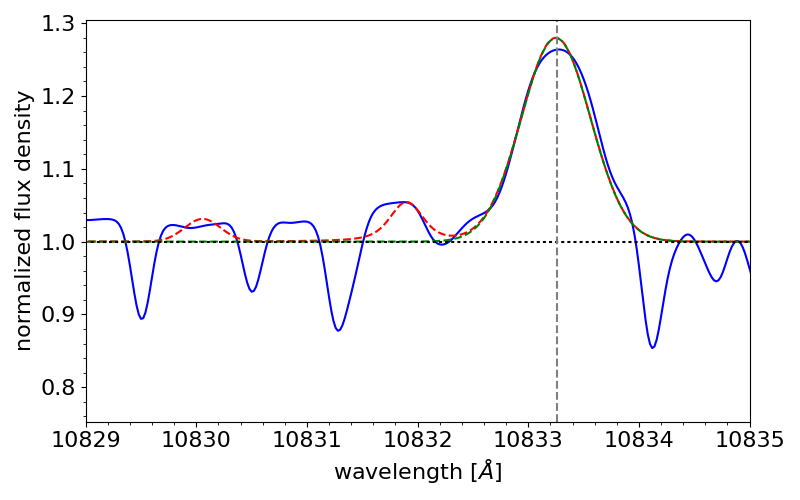}
      \caption{Fitting Voigt profiles in the wavelength range around the \ion{He}{i}~IR triplet
      (10\,829--10\,835\,$\AA$). 
      Three models (solid blue lines) at different states of activity are shown: a model with \ion{He}{i}~IR 
      in absorption (\textit{top panel}, number 042 from \citet{Hintz2019A&A...623A.136H}) or 
      number A2 from the model series A, details are given in Sect.~\ref{Models_new}), one nearly filled in 
      (\textit{middle panel}, number 149), and one with the line in emission (\textit{bottom panel}, number 121). 
      The red dashed line is the best fit of four Voigt profiles in the wavelength range of 10\,829--10\,835$\,\AA$. 
      The dashed green line represents the single Voigt profile fit to the center of the two 
      indistinguishable red components of the \ion{He}{i}~IR triplet at 10\,833.25$\,\AA$ (dashed vertical line). 
      The pEW of the \ion{He}{i}~IR line is calculated from the green Voigt profile. 
              }
         \label{pew_he10830_fits_abs_em_trans}
   \end{figure}

   In the following, we give a brief overview of the method and refer to 
   \citet{Fuhrmeister2019A&A...632A..24F} for more details.
   We used an empirical function consisting of four Voigt profiles to approximate the spectrum in 
   the wavelength range of 10\,829--10\,835$\,\AA$. The Voigt components represent the relevant spectral
   lines,
   including the \ion{He}{i}~IR line. 
   The two red components of the \ion{He}{i}~IR triplet at 10\,833.217 and 10\,833.306$\,\AA$ are unresolved, and, 
   therefore, represented by one Voigt profile centered at 10\,833.25$\,\AA$, which is used to
   estimate the pEW. 
   The weak blue triplet component is neglected in the calculations of this pEW 
   because it is blended with an unidentified absorption line. 
   Figure~\ref{pew_he10830_fits_abs_em_trans} shows the Voigt profile fits of the \ion{He}{i}~IR line 
   for three model spectra of different states of activity. 
   While the fit of the more or less filled-in line (middle panel) hardly reproduces the shape of the line, 
   clear absorption and emission lines (top and bottom panels, respectively) 
   are well fit by the respective Voigt profiles. Therefore, 
   pEW values close to zero may be problematic. 
   
   To determine the pEW of 
   the H$\alpha$, \ion{He}{i}~D$_3$, 
   \ion{Na}{i}~D$_2$, and the bluest \ion{Ca}{ii}~IRT line, 
   we integrated the model spectrum in the respective 
   wavelength ranges following \citet{Fuhrmeister2019A&A...632A..24F} for the stellar sample. 
   The central wavelengths of the lines, the line widths, and the ranges of the reference bands 
   are adopted from \citet{Fuhrmeister2019A&A...632A..24F}.

\subsection{Previous chromospheric models} \label{Models}
   
   The stellar atmosphere code 
   PHOENIX\footnote{\url{https://www.physik.uni-hamburg.de/en/hs/group-hauschildt/research/phoenix.html}} 
   has been developed 
   for calculations of atmospheres and spectra 
   from various objects such as stars, planets, novae, and supernovae 
   \citep{Hauschildt1992JQSRT..47..433H, Hauschildt1993JQSRT..50..301H, Hauschildt1999JCoAM.109...41H}. 
   A recent library of PHOENIX model photospheres was calculated by \citet{Husser2013} 
   in the effective temperature range of 2\,300--12\,000\,K, 
   covering a wide range of different stellar spectral types.
   Based on one of these photospheric models, \citet{Hintz2019A&A...623A.136H} computed chromosphere models for M dwarfs using the PHOENIX code, but 
   extending the photospheric PHOENIX model, which was computed by 
   \citet{Husser2013} with $T_{\rm eff} = 3500\,$K, $\log g=5.0\,$dex, 
   $[\mathrm{Fe}/ \mathrm{H}] = 0.0\,\mathrm{dex,}$ and $[\alpha/\mathrm{Fe}] = 0.0\,\mathrm{dex,}$ 
   by a chromosphere and transition region, assuming a temperature structure characterized 
   by linear sections in the logarithm of the column mass density 
   for the lower and upper chromosphere as well as for the transition region. 
   In total, we varied six free parameters in the temperature structure of the chromosphere and 
   computed a set of 166 one-dimensional spherically symmetric chromospheric models. 
   In creating the models, the semi-empirical, solar VAL~C temperature structure
   was parameterized and adjusted to the M-dwarf sample. 
   The parameterized ad hoc temperature structure represents the unknown heating mechanisms in the upper atmosphere. 
   In addition, we assumed hydrostatic equilibrium and 
   neglected any acoustic and magnetic waves. 
   Convergence was achieved  
   using nonlocal thermodynamic equilibrium (NLTE) calculations for the chromospheric lines. 
   The grid encompasses 
   models of various levels of activity. 
   The spectra of the inactive models typically show the chromospheric lines in absorption, 
   while with increasing activity, the lines go into emission. \citet{Hintz2019A&A...623A.136H} optimized their models to simultaneously fit the 
   \ion{Na}{i}~D$_2$, H$\alpha$, and the bluest \ion{Ca}{ii}~IRT line 
   of the stellar sample described above. 
   
\subsection{Construction of new chromospheric models} \label{Models_new}

   \begin{table}[t]
   \caption{Parameters of the extended model series A and B based on models 
   042 and 080 from \citet{Hintz2019A&A...623A.136H}, 
   hereafter A2 and B4 in the new model series, respectively. }    \vspace{-2em}         
   \label{model_extension}      
   \begin{center}
   \footnotesize
   \begin{tabular}{l c c c c c c c}
   \hline\hline       
Model & $m_\mathrm{min}^{\,\,a}$ & $m_\mathrm{mid}^{\,\,a}$ & $T_\mathrm{mid}^{\,\,a}$ & $m_\mathrm{top}^{\,\,a}$ & $T_\mathrm{top}^{\,\,a}$ & $\nabla_{\rm TR}^{\,\,a}$ & $m_i / m_0 ^{\,\,b}$ \\ 
      & $[\mathrm{dex}]$ & $[\mathrm{dex}]$ & $[\mathrm{K}]$ & $[\mathrm{dex}]$ & $[\mathrm{K}]$ & $[\mathrm{dex}]$ &  \\ 
 \hline 
A1   & -3.0   & -4.1   & 5500   & -5.5   & 6500   & 7.5   & 0.37 \\ 
A2   & -2.5   & -3.6   & 5500   & -5.0   & 6500   & 7.5   & 1.0   \\ 
A3   & -2.0   & -3.1   & 5500   & -4.5   & 6500   & 7.5   & 3.16 \\ 
A4   & -1.9  & -3.0  & 5500   & -4.4  & 6500   & 7.5   & 4.47 \\ 
 \hline 
B1   & -2.5   & -3.5   & 5500   & -5.0   & 7500   & 8.5   &  0.1   \\ 
B2   & -2.3   & -3.3   & 5500   & -4.8   & 7500   & 8.5   &  0.16 \\ 
B3   & -1.8   & -2.8   & 5500   & -4.3   & 7500   & 8.5   &  0.50 \\ 
B4   & -1.5   & -2.5   & 5500   & -4.0   & 7500   & 8.5   &  1.0   \\ 
B5   & -1.3   & -2.3   & 5500   & -3.8   & 7500   & 8.5   &  1.59 \\ 
B6   & -1.2  & -2.2  & 5500   & -3.7  & 7500   & 8.5   &  2.24 \\ 
   \hline   
   \end{tabular}
   \end{center}
   \footnotesize{$^{a}$
   The six parameters are consistent with the definitions of \citet{Hintz2019A&A...623A.136H}. 
   The column mass density $m_\mathrm{min}$ represents the onset of the lower chromosphere 
   and also marks the location of the temperature minimum. The end of the lower chromosphere 
   is located at the column mass density $m_\mathrm{mid}$ and the temperature $T_\mathrm{mid}$. 
   The end of the upper chromosphere and onset of the transition region is located at 
   ($m_\mathrm{top}$, $T_\mathrm{top}$), and $\nabla_{\rm TR}$ gives the temperature gradient 
   of the transition region. \\
   $^{b}$ The parameter $m_i / m_0$ defines the mass load in comparison to the respective basis model.
   }
   \end{table}
   
   \begin{figure}[t]
   \includegraphics[width=0.5\textwidth]{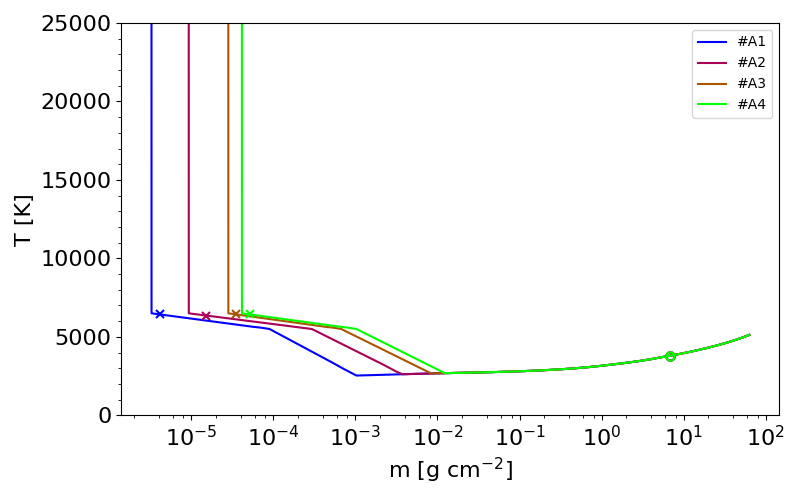}
   \includegraphics[width=0.5\textwidth]{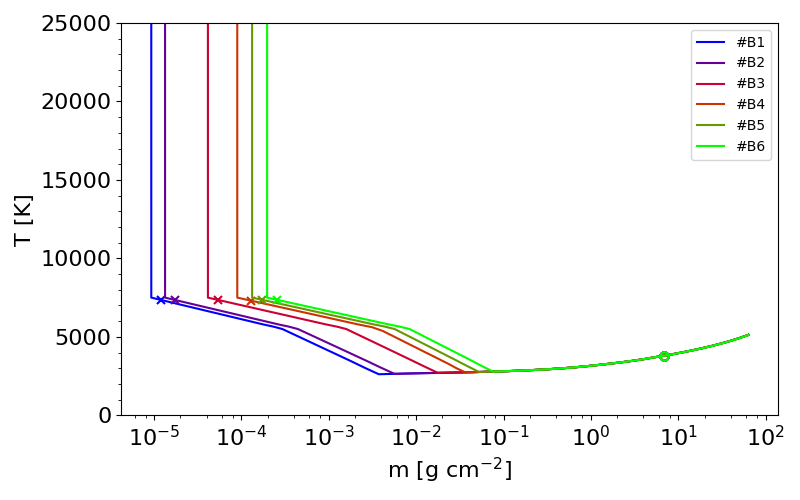}
   \caption{
   Model temperature profiles as a function of the column mass densities. 
   The respective model parameters are 
   given in Table~\ref{model_extension}. 
   The temperature structures of series A are shown in the \textit{top panel}, and 
   those of series B are depicted in the \textit{bottom panel}. 
   Models A2 and B4 in these series correspond to models 
   042 and 080 from \citet{Hintz2019A&A...623A.136H}, respectively. 
   In both series, we also mark the structural positions 
   where the \ion{He}{i}~IR line core (crosses) and continuum at $\lambda = 10\,834.5\,\AA$ (open circles) 
   approach an optical depth of $\tau=1$, i.\,e., where both become optically thick.
              }
         \label{T_struct1}
   \end{figure}
   
   Here we extend our previous study that we performed in \citet{Hintz2019A&A...623A.136H} by constructing two
   series of models with the specific goal of 
   studying the behavior of the \ion{He}{i}~IR line in terms of the activity state. 
   We varied the activity level of a given chromospheric temperature structure 
   by shifting the whole density structure. Models shifted farther inward toward higher densities 
   correspond to higher activity states. 
   In particular, we performed systematic density shifts for two 
   inactive best-fit models of \citet{Hintz2019A&A...623A.136H}, 
   those designated numbers 042 and 080 therein.

   The temperature structure of these two models was shifted toward higher and lower densities 
   following the approach of \citet{Andretta1997ApJ...489..375A}. 
   Table~\ref{model_extension} lists the model parameters, and 
   Fig.~\ref{T_struct1} depicts the temperature structures of these series of models. 
   The change in the column mass density is described by $m_i / m_{0}$ 
   relative to the basis model 042, 
   hereafter called model A2 within the new series designated by A, while former model
   080 is hereafter called model B4 in series B. 
   Model A1 (blue line in Fig.~\ref{T_struct1}) of the A series represents the least active state with 
   the column mass density of the layers being about ten times lower than that of the original model A2. 
   Model A4 (green line) has the largest inward shift ($m_i / m_0 = 4.47$), corresponding to the
   highest activity state of the series. 
   
   Because we shifted the atmospheric structure on a given discontinuous column mass grid, 
   we obtained deviations in the gradients of the different linear sections. 
   We allowed a tolerance deviation at maximum of 15\,\% for the gradients compared to 
   the gradients of the respective basis model. 
   Therefore the series are not equidistant on the column mass grid. 
   While the NLTE level occupation numbers of the models in series A and B converge 
   during the iteration process of the model computation as described in \citet{Hintz2019A&A...623A.136H}, 
   further inward shifts of the innermost models come along with nonconverged atmospheres. 
   Thus, models A4 and B6 yield the limit 
   of the inward density shift of the respective series. 
      
\subsection{New model spectra} \label{Models_new_spectra}

   \begin{figure*}[ht]
   \centering
   \includegraphics[width=0.49\textwidth]{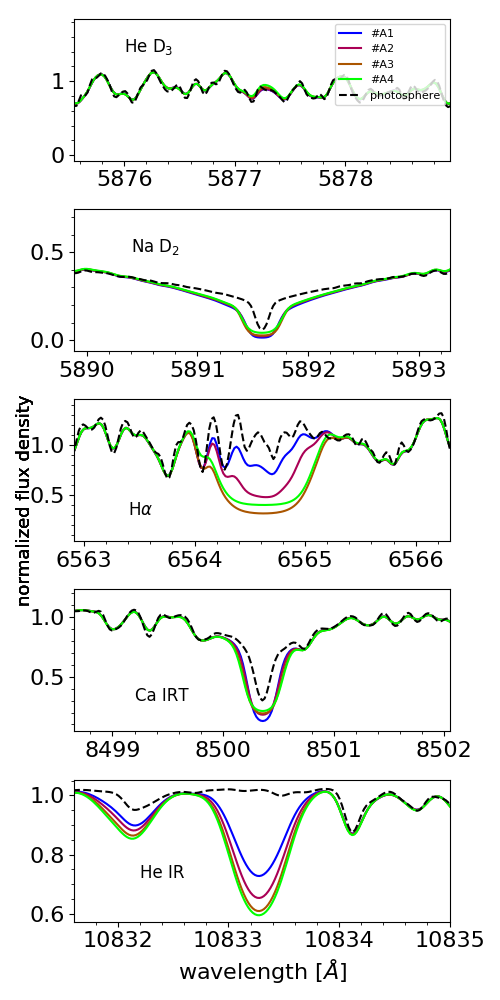}
   \includegraphics[width=0.49\textwidth]{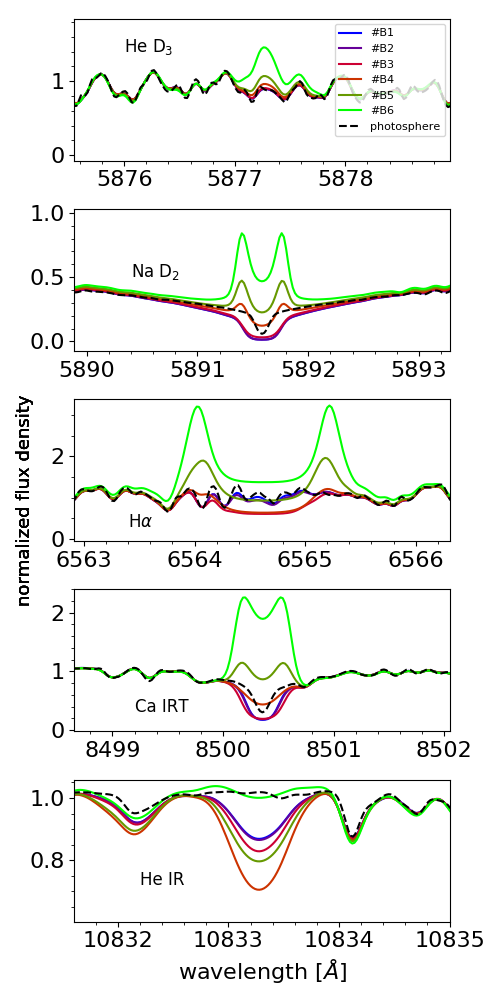}
   \caption{
   Spectra according to the temperature structures of the new series 
   A (\textit{left panel}) and B (\textit{right panel}) of models in 
   Fig.~\ref{T_struct1} (corresponding color-coding) 
   in the spectral ranges of \ion{He}{i}~D$_3$, \ion{Na}{i}~D$_2$, H$\alpha$, 
   the bluest \ion{Ca}{ii}~IRT line, and the \ion{He}{i}~IR line (from \textit{top} to \textit{bottom}). 
   For comparison, the underlying photosphere model ($T_{\rm eff} = 3500\,$K, $\log g = 5.0\,\mathrm{dex}$, 
   $[\mathrm{Fe}/ \mathrm{H}] = 0.0\,\mathrm{dex}$, $[\alpha/\mathrm{Fe}] = 0.0\,\mathrm{dex}$) 
   is plotted as well. 
   The photosphere is taken from \citet{Husser2013}. 
   }
         \label{series_new_chroms}
   \end{figure*}
   
   From both new series A and B of model atmospheres as given by the chromospheric temperature structures 
   in Fig.~\ref{T_struct1} we obtained model spectra that are shown in the wavelength ranges of 
   the chromospheric lines of \ion{He}{i}~D$_3$, \ion{Na}{i}~D$_2$, H$\alpha$, the bluest \ion{Ca}{ii}~IRT line, 
   and the \ion{He}{i}~IR line in Fig.~\ref{series_new_chroms}. 
   The series show the evolution of these chromospheric lines with increasing activity, 
   that is, the inward shifting of the models. 
   The outermost models with the respective lowest densities, A1 and B1, already show absorption in the \ion{He}{i}~IR line. 
   Models representing higher activity levels
   lead to an increase in absorption strength, with 
   model A4 exhibiting the strongest absorption line in  series A. 
   In series B, model B4 displays the highest absorption 
   depth, while  
   for models B5 and B6, the \ion{He}{i}~IR line fills in. 
   In Fig.~\ref{T_struct1} we also indicate where the \ion{He}{i}~IR line core 
   and neighboring continuum become optically thick. 
   The corresponding temperatures for the \ion{He}{i}~IR line core 
   range between 6350\,K and 6450\,K for series A and 
   between 7300\,K and 7400\,K for series B. 
   The continuum temperatures are in both cases at $\sim 3800\,$K. 
   The optically thick temperature regions of the line core indicate 
   a \ion{He}{i}~IR line formation in the upper chromosphere. 
   Furthermore, we determined whether 
   geometric effects caused by
   the chromospheric extension above the photosphere contribute
   to the \ion{He}{i}~IR line formation in our spherically symmetric models. 
   We were unable to find significant geometric effects 
   with respect to the \ion{He}{i}~IR line formation. 
      
   A comparable behavior is observed in the H$\alpha$ line, where model A4 
   already shows fill-in. 
   This means that the evolution of H$\alpha$ seems to be shifted compared to that of the \ion{He}{i}~IR line
   in series A. 
   In series B, model B3 displays maximum absorption in H$\alpha$ and model B4 just 
   starts to fill in, which is slightly more strongly visible in the wings. 
   In models B5 and B6, the H$\alpha$ line tends to go into emission. 
   However, the H$\alpha$ lines of models B5 and especially B6 exhibit strong 
   trough-like self-absorption features, suggesting that at least model B6 
   does not represent a real stellar spectrum because even during flares no such broad self-absorption troughs
   have been observed. This demonstrates that not every parameterized chromosphere leads
   to realistic spectra. 
   
   The \ion{He}{i}~D$_3$ line is not visible at all in series A. 
   Only in series B does the line arise beyond the continuum when the activity levels increase: 
   model B5 shows the line in slight emission, and a clear \ion{He}{i}~D$_3$ emission line appears in model B6. 
   In the cases of the \ion{Na}{i}~D$_2$ and the bluest \ion{Ca}{ii}~IRT line, the evolution is 
   different than in the other cases: the lines start at a maximum absorption depth and fill in for both series 
   while shifting the whole atmospheric structure farther inward. 
   For series B, the response of the lines is again stronger, and for the two innermost models B5 and B6, 
   these lines go into emission.

\subsection{Flare classification and model selection} \label{Models_flares}
   
   In our investigation of the formation of \ion{He}{i}~IR line, we 
   focused on quiet atmospheres. Therefore
   we identified models representing high-activity states (flaring). 
   To that end, 
   we followed observational findings in the spectra. 
   Because H$\alpha$ can be observed in emission during the quiescent state, 
   the line is not well suited as a flare criterion; the same applies for most chromospheric
   lines in the optical. In contrast, we are not aware of Paschen emission line observations
   outside of flares, while the line was observed in emission during M-dwarf flares 
   \citep{Schmidt2012ApJ...745...14S, Fuhrmeister2008A&A...487..293F, Liebert1999ApJ...519..345L, Paulson2006PASP..118..227P}. 
   Therefore we used the Pa$\beta$ emission as a discriminator for flaring activity. 
   With this criterion, we identified the highest activity states. 
   
   In particular, we computed the pEW for Pa$\beta$ at $12821.57\,\AA$. 
   For the line core, we assumed a width of $0.5\,\AA,$ and 
   the reference bands were located at $12\,811.5 \pm 1.0\,\AA$ and $12\,826 \pm 1.0\,\AA$. 
   Each model yielding a pEW value below $-0.02\,\AA$ 
   was considered a flare model 
   and was omitted in the detailed analysis of the \ion{He}{i}~IR line. 
   One model \citep[number 166 from ][]{Hintz2019A&A...623A.136H} passed this criterion, 
   but the Pa$\beta$ line developed broad emission wings that reached into the reference bands. 
   Including this model, a total of
   81 models were flagged as flaring and were excluded from further consideration. 
   
   We also neglected models with a transition region 
   onset below $6000\,$K. It might be argued that the onset of the transition region occurs where
   hydrogen becomes completely ionized and hence no longer is an efficient cooling agent. 
   While this occurs only at about 8000\,K, hydrogen becomes partially ionized above $5000\,$K 
   so that cooling starts to diminish \citep{Ayres1979ApJ...228..509A}. 
   
   In total, we investigated a subsample of 58 models from \citet{Hintz2019A&A...623A.136H} 
   that fulfilled our selection criteria for inactive chromospheres; 
   these models and the parameter configurations are listed in 
   Table~\ref{table_model_grid}.  
   Because we are interested in investigating nonflaring models 
   and the evolution of the \ion{He}{i}~IR line as a function of increasing activity, 
   we took two best-fit models of inactive stars from 
   our previous work and varied their activity state. 
   We  thus included eight newly 
   calculated models from series A and B in our investigation even though 
   models B5 and B6 did not pass the Pa$\beta$ criterion. 
   
\section{Results and discussion}

\subsection{Model EUV flux and its effect on the \ion{He}{i}~IR line} \label{EUV}

   The wavelength range of the EUV 
   is self-consistently calculated within the PHOENIX models 
   according to the prescribed atmospheric structure. The 
   maximum temperature is 98\,000\,K in all of our models. 
   We did not extend the models to 
   higher temperatures because then conduction plays a major
   role that is not incorporated in the PHOENIX models. 
   In this regard,
   the models underpredict the EUV radiation field because the upper part of the transition region
   and the corona are neglected. 
   
\begin{figure}[t]
   \centering
   \includegraphics[width=0.5\textwidth]{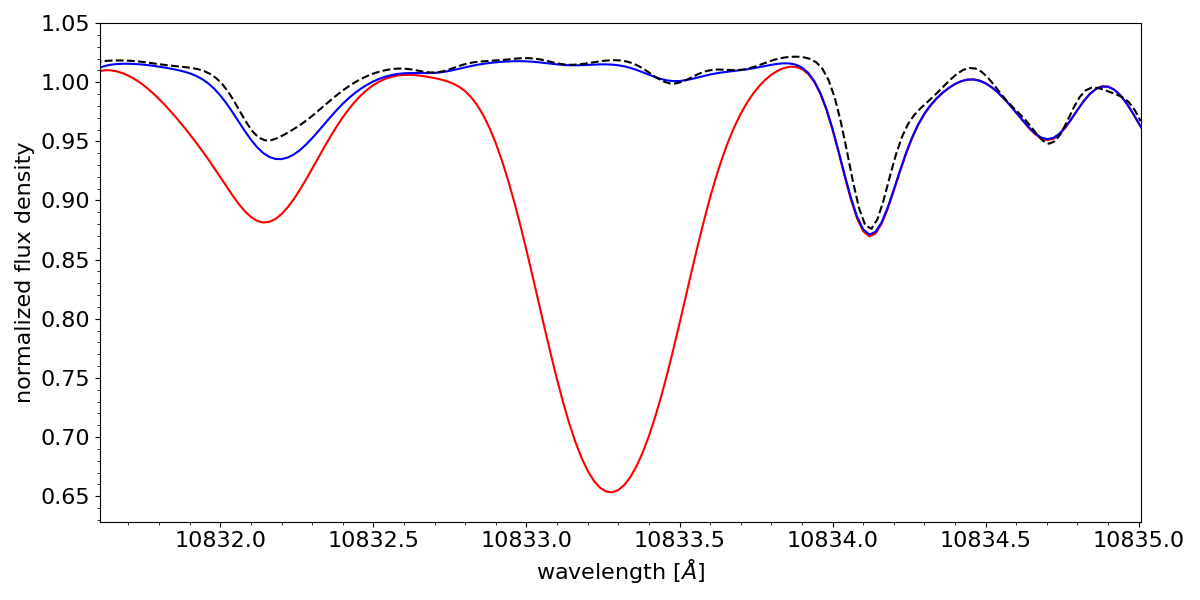} 
   \includegraphics[width=0.5\textwidth]{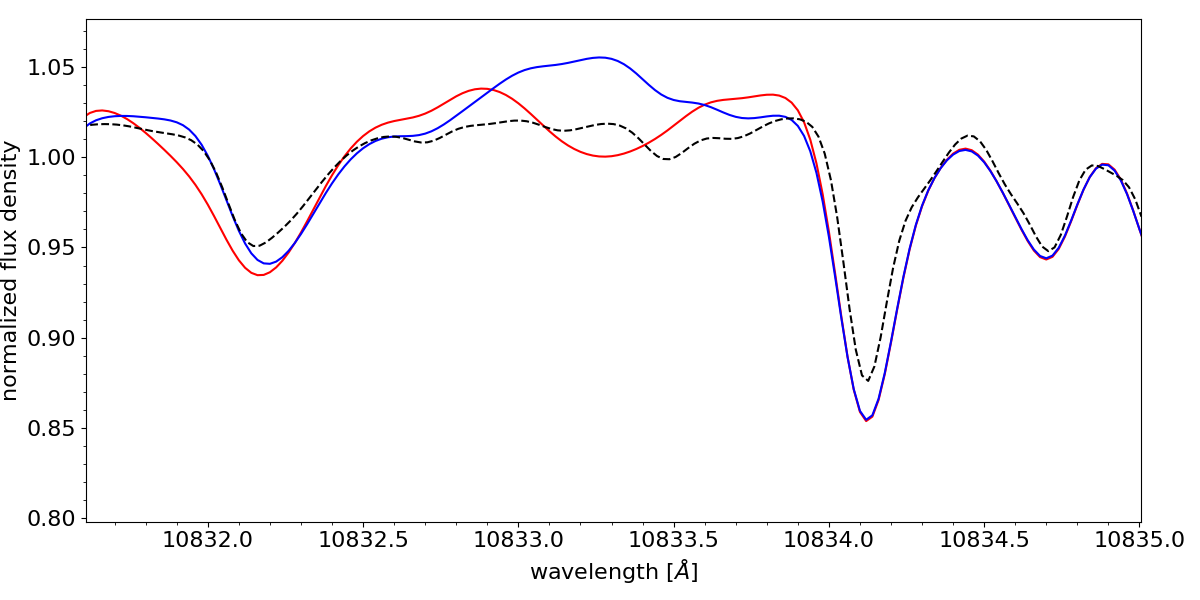} 
   \caption{
   Spectral ranges of the \ion{He}{i}~IR line in 
   models A2 (\textit{upper panel}) and B6 (\textit{lower panel})  
   in the original configuration (red lines) and when the radiation field below $504\,\AA$ (blue lines) is omitted. 
   The dashed black line shows the underlying photosphere alone (from \citet{Husser2013} with 
   $T_{\rm eff} = 3500\,$K, $\log g=5.0\,$dex, 
   $[\mathrm{Fe}/ \mathrm{H}] = 0.0\,\mathrm{dex}$, and 
   $[\alpha/\mathrm{Fe}] = 0.0\,\mathrm{dex}$ ). 
   }
         \label{no_tr}
   \end{figure}

   \citet{Andretta1997ApJ...489..375A} highlighted that it is important that the EUV 
   radiation field is irradiated from the solar transition region and corona 
   in the formation of the \ion{He}{i} spectrum of the Sun. 
   Omitting the radiation field in the VAL~C model of the quiet Sun, the authors obtained 
   a model spectrum without any visible \ion{He}{i} line. 
   We repeated this exercise exemplarily for one of our inactive models A2 and  of our most active
   model B6, and show the resulting \ion{He}{i} IR line in Fig.~\ref{no_tr}. 
   While the original model spectra were calculated down to a wavelength of $10\,\AA$, 
   we cut off the wavelength range $\lambda \le 504\,\AA$ and left 
   everything else unchanged. 
   In the spectrum of the modified model A2, the \ion{He}{i}~IR line 
   completely vanishes, while the self absorption of the fill-in seen in model B6 
   turns into a slight emission line without any sign of self-absorption. 
   This indicates that in the PHOENIX model calculations the PR mechanism is 
   important in populating the metastable $2\,^3S$ level of helium 
   to obtain any absorption feature in the first place. 
   Because the PR mechanism is eliminated in the modified model B6, 
   the emission is caused by collisional excitation alone. 
   
   \begin{figure}[t]
   \centering
   \includegraphics[width=0.5\textwidth]{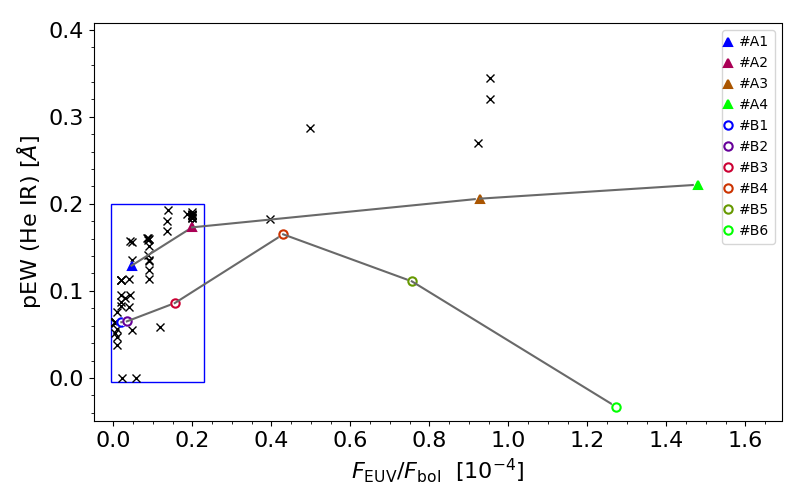}
   \includegraphics[width=0.5\textwidth]{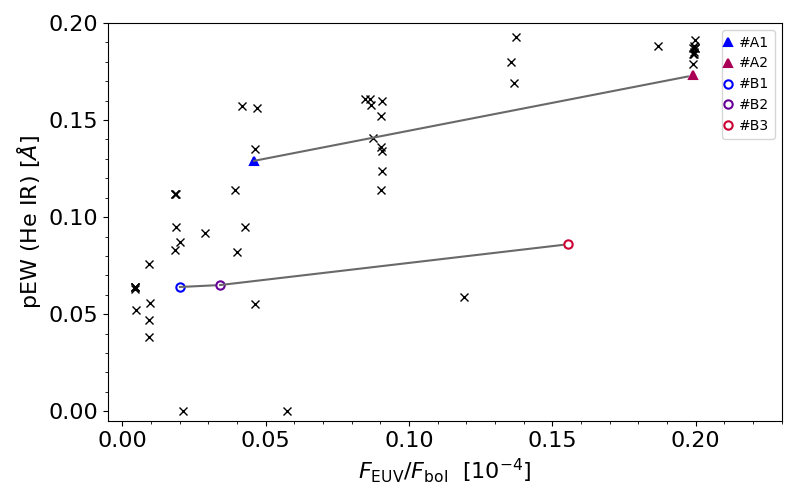}
   \caption{\ion{He}{i}~IR pEW as a function of the 
      integrated EUV flux at $\lambda \le 504\, \AA$ (vacuum). 
      The color-coded models correspond to the new series A and B of models 
      as given in Fig.\ref{T_struct1}. 
      Black crosses represent the previously nonflaring models from \citet{Hintz2019A&A...623A.136H} 
      as given in Table~\ref{table_model_grid}. 
      The region marked by the blue rectangular in the \textit{upper panel} is 
      enlarged in the \textit{lower panel} and shows our most quiescent models. 
              }
         \label{pEW_EUV}
   \end{figure}
   
   We calculated the emanating EUV radiation of our PHOENIX models by integrating the flux density 
   in the wavelength range of $10\,\AA \le \lambda \le 504\,\AA$ ($F_{\rm EUV}$) and 
   related it to the integrated bolometric flux ($F_{\rm bol}$). 
   In Fig.~\ref{pEW_EUV} we show the pEW(\ion{He}{i}~IR) as a function of the 
   EUV radiation field as measured by $F_{\rm EUV}/F_{\rm bol}$. 
   The values of $F_{\rm EUV}/F_{\rm bol}$ from our nonflaring PHOENIX models 
   (all models investigated in this work except for models B5 and B6) 
   as shown in Fig.~\ref{pEW_EUV}
   range from $4.5 \times 10^{-7}$ for model 047 to 
   $1.48 \times 10^{-4}$ for model A4. 
   These values are difficult to compare to observations because EUV observations for M dwarfs are especially hard to come by.
   Nevertheless, we can compare our values to the 
   estimated EUV radiation field of GJ~176 from \citet{Loyd2016ApJ...824..102L}, 
   obtained in the context of the Measurements of the Ultraviolet Spectral 
   Characteristics of Low-mass Exoplanetary Systems (MUSCLES) Treasury Survey. 
   The estimated $F_{\rm EUV}/F_{\rm bol}$ value of GJ~176 from \citet{Loyd2016ApJ...824..102L} is 
   in the order of $10^{-5}$, which is slightly higher than the typical value that 
   we obtain for our nonflaring PHOENIX models; the median of $F_{\rm EUV}/F_{\rm bol}$ 
   of the nonflaring models is $8.9 \times 10^{-6}$. 
   GJ~176 is slightly above our effective temperature
   range with $T_{\rm eff} = 3689\pm54\,$K \citep[][]{Passegger2019A&A...627A.161P}. 
   We therefore conclude that the calculated $F_{\rm EUV}/F_{\rm bol}$ values from 
   our nonflaring PHOENIX chromospheres are compatible with observational results of EUV radiation, 
   although our models neglect the 
   temperature region of the upper transition and corona because of the 
   model temperature limit of 98\,000\,K. 
   However, compared to real stars, our models presumably contain too much material 
   in the lower transition region, and this  might therefore compensate for the material 
   that we omitted in the upper transition region and corona. 
   This means that we obtain realistic $F_{\rm EUV}/F_{\rm bol}$ within our models 
   despite this inadequacy. 
   For a recent detailed analysis of a synthetic EUV PHOENIX spectrum computed for 
   the inactive M~dwarf Ross~905, we refer to the study by 
   \citet[][]{Peacock2019ApJ...886...77P}. 
   
   The graph in Fig.~\ref{pEW_EUV} shows that an increase in 
   the strength of the radiation field in the transition 
   region tends to deepen the \ion{He}{i}~IR line. 
   The effect of the radiation field 
   on the line strength decreases at higher irradiation levels. 
   For $F_{\rm EUV}/F_{\rm bol} \lesssim 2.0 \times 10^{-5}$, 
   absorption in the \ion{He}{i}~IR line strongly increases in response to
   a comparably weak rise of the integrated EUV flux. 
   When the models are shifted farther inward, the EUV flux increases more strongly, but the \ion{He}{i}~IR 
   line strength only changes slowly. 
   The pEW(\ion{He}{i}~IR) seems to saturate or even starts to fill in for higher $F_{\rm EUV}/F_{\rm bol}$ values. 
   Our modeling yields a
   maximum absorption depth at ${\rm pEW} = 0.35\,\AA$ 
   for model 057. 
      
   \begin{figure*}[t!]
   \centering
   \includegraphics[width=0.49\textwidth]{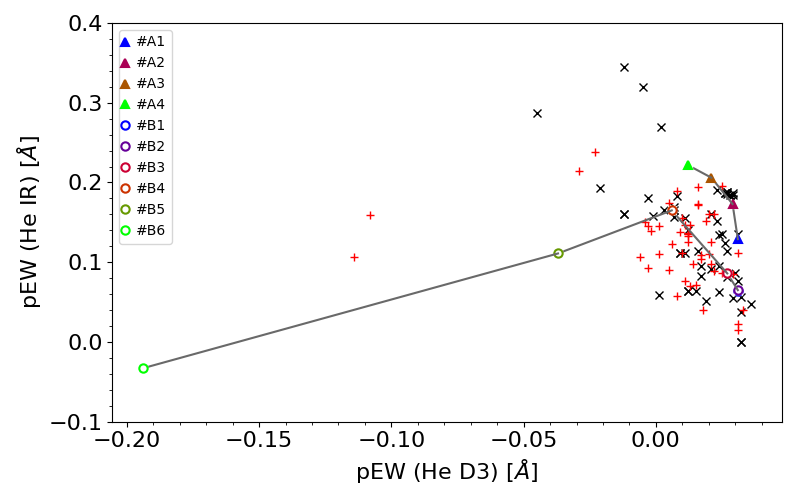}
   \includegraphics[width=0.49\textwidth]{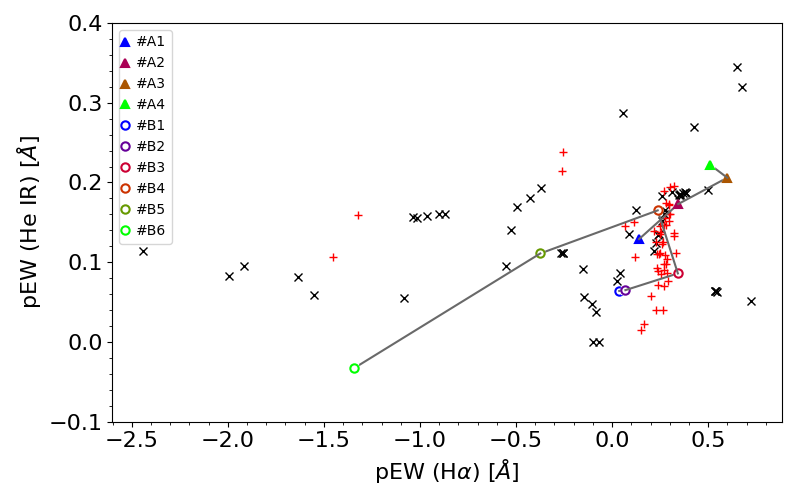}
      \caption{pEWs of the \ion{He}{i}~IR line as a function of the pEWs of the \ion{He}{i}~D$_3$ (\textit{left panel}), 
      and H$\alpha$ (\textit{right panel}) lines. 
      Black crosses represent the nonflaring models from \citet{Hintz2019A&A...623A.136H}. 
      Series A and B 
      are depicted by color-coded squares 
      (the color-coding corresponds to that of Fig.~\ref{T_struct1}). 
      CARMENES observations of the investigated M2--M3\,V stars are flagged by red pluses. 
      In the observed pEW(\ion{He}{i}~D$_3$) of the stellar sample, 
      we add an offset of $0.06\,\AA$ because these measurements are influenced by a decline in flux 
      in the respective normalization reference bands from the blue to the red continuum. 
              }
         \label{pew_he10830_vs_others}
   \end{figure*}
   
   The comparison of the results for series A and B shows that both series 
   yield a qualitatively similar behavior within the low-activity regime of 
   the pEW(\ion{He}{i}~IR) vs. $F_{\rm EUV}/F_{\rm bol}$ plane, although 
   they differ in their model parameters such as the temperature gradient in the transition region. 
   However, the extent of the response of the radiation field and the line depth 
   on the shift of the models depends on the series. 
   In both cases, the chromospheres are equal in thickness on the column mass density scale, 
   but the gradients of the upper chromospheres and 
   transition regions differ from each other. 
   This difference affects the density at the temperatures where the PR mechanism works in the \ion{He}{i}~IR line formation. 
   Fig.~\ref{pEW_EUV} shows that the effect of shifting the prescribed atmospheric structure 
   in series B is stronger than in A. 
   In model series A, the line absorption monotonically 
   increases with increasing $F_{\rm EUV}/F_{\rm bol}$. 
   In the other model series, model B4 exhibits the highest absorption level (${\rm pEW} = 0.165\,\AA$) at $F_{\rm EUV}/F_{\rm bol} = 4.3 \times 10^{-5}$. 
   A further inward shift results in filling in the line with a further increase in the 
   radiation field. 
   Because the maximum absorption depth is not reached for series A, the
   saturation level does not appear to be determined solely by the position of the chromospheric structure 
   in the atmosphere, but also depends on other parameters such as temperature gradients or the temperature at the
   onset of the transition region. 
   
   \subsection{Comparison to observations}
   
   \citet{Fuhrmeister2019A&A...632A..24F} compared the pEW(\ion{He}{i}~IR) measurements 
   of their investigated M-dwarf sample to X-ray observations that can serve 
   as a proxy of the EUV radiation field. 
   For their inactive and quiescent stars, they found that for low $\log L_{\rm X} / L_{\rm bol}$ values  the full range
   of pEW(\ion{He}{i}~IR) values was observed. For high
   X-ray luminosities, only high pEW(\ion{He}{i}~IR) values were seen, that is, there is some 
   sort of lower envelope that is correlated with $\log L_{\rm X} / L_{\rm bol}$. 
   This lower envelope has the steepest gradient for M0 dwarfs, while later subtypes exhibit lower gradients, 
   with M3--4 dwarfs showing no dependence on $\log L_{\rm X} / L_{\rm bol}$. 
   For the M2--3 dwarfs considered here, they therefore only  found a very weak tendency for high 
   $\log L_{\rm X} / L_{\rm bol}$ values excluding low pEW(\ion{He}{i}~IR) values. 
   
   Our models, on the other hand, show a correlation between the 
   pEW(\ion{He}{i}~IR) and $F_{\rm EUV}/F_{\rm bol}$ measurements. Nevertheless, this need not necessarily be 
   a discrepancy. We do not know which $F_{\rm EUV}/F_{\rm bol}$ values the 
   observations span, but when we take only the nonflaring measurements 
   into account, they are probably covered by a subset of models from the 
   zoom-in (lower panel in Fig.~\ref{pEW_EUV}). When only the models with  
   $0.04\times 10^{-4} < F_{\rm EUV}/F_{\rm bol} < 0.1\times 10^{-4}$ 
   are considered, the situation starts to look similar to 
   the observations: while the upper envelope of the pEW(\ion{He}{i}~IR) values is 
   quite constant, the lower envelope shows some rise. Moreover, the spread 
   in pEW(\ion{He}{i}~IR) seen for the models for a distinct $F_{\rm EUV}/F_{\rm bol}$ value should 
   dilute a possible correlation even more in the observations because they are 
   additionally prone to measurement errors, errors arising from using X-ray 
   fluxes as approximation for $F_{\rm EUV}$, and from nonsimultaneous measurements 
   in X-ray and pEW(\ion{He}{i}~IR). 
   
   Nevertheless, \citet{Fuhrmeister2019A&A...632A..24F} also found some saturation limit that depended on spectral type, which is about pEW$_{\rm sat}$=0.25\,\AA for spectral types M2.0--2.5\,V and 
   pEW$_{\rm sat}$=0.2\,\AA\, for M3.0--3.5\,V stars. This agrees with the overall saturation limit 
   found by the models, although individual models predict even deeper \ion{He}{i}~IR lines.     
   The EUV radiation where the response of the \ion{He}{i}~IR line saturates in our model 
   series B may be connected with a change in formation mechanism, that is, 
   the collisional excitation becomes dominant because the models B5 and B6 are flare models and the observations by 
   \citet{Fuhrmeister2019A&A...632A..24F} with positive pEW(\ion{He}{i}~IR) are 
   connected with flaring events. 
   Omitting the EUV radiation in the models also supports this conclusion because an inactive chromosphere model without EUV radiation does not show any signal of the \ion{He}{i}~IR line, 
   and an active model tends to show the \ion{He}{i}~IR line in emission.

   \begin{figure*}[t!]
   \centering
   \includegraphics[width=0.9\textwidth]{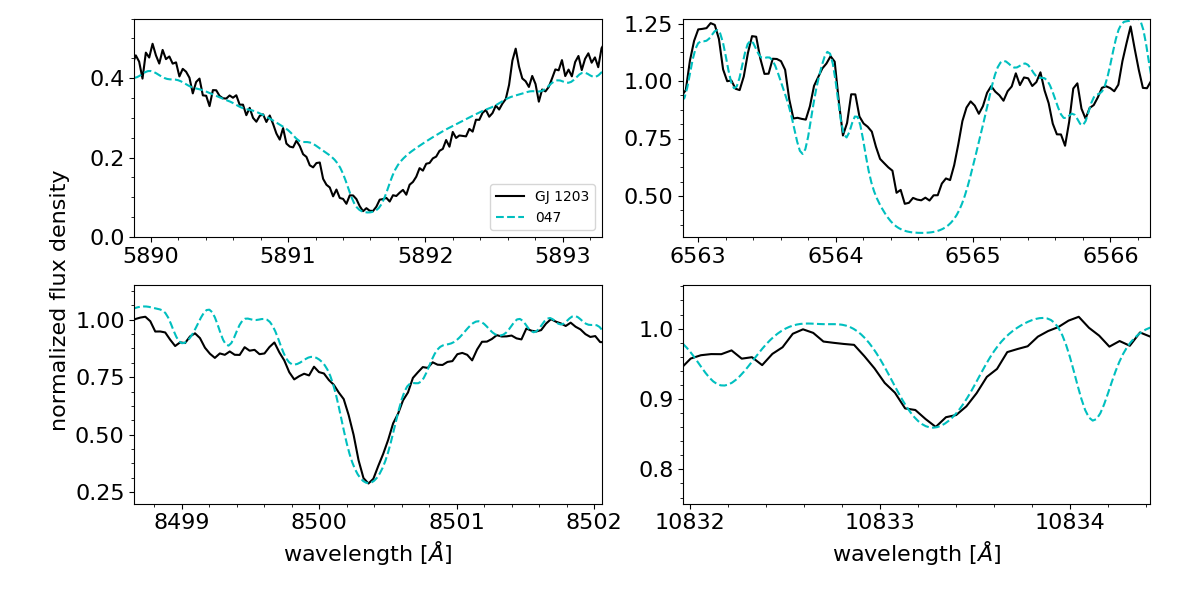}
   \includegraphics[width=0.9\textwidth]{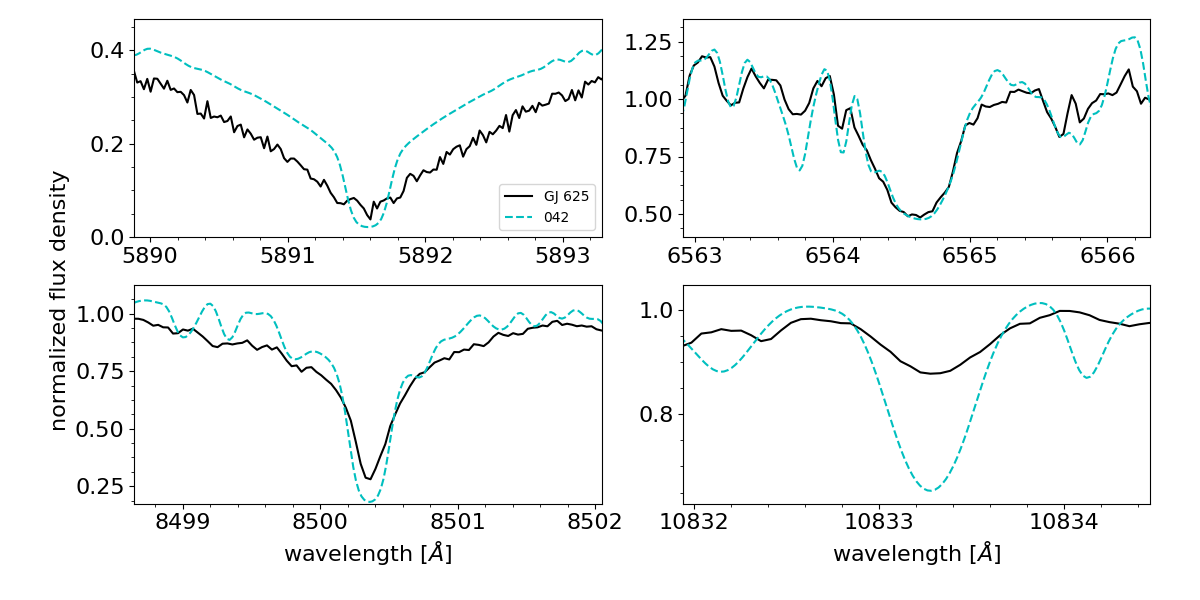}
      \caption{
               \textit{Upper panel:} 
               Comparison of the best-fit model spectrum (dashed cyan line, model 047) 
               to an observed spectrum (solid black line) of GJ~1203 
               optimized for the lines of \ion{Na}{i}~D$_2$ (\textit{top left panel}), 
               H$\alpha$ (\textit{top right panel}), and the bluest \ion{Ca}{ii}~IRT line  (\textit{bottom left panel}). 
               In the range of the \ion{He}{i}~IR line (\textit{bottom right panel}), the 
               observed CARMENES template obtained with SERVAL is shown. 
               The emission line redward of the \ion{Na}{i}~D$_2$ line in the observation of GJ~1203 
               is an airglow line. 
               \textit{Bottom panel:} 
               Same as in the upper plot for GJ~625 (model 042). 
               }
         \label{active_obs_mod_gj70}
   \end{figure*}

\subsection{\ion{He}{i}~IR in relation to other chromospheric lines} \label{best-fits}   
   In Fig.~\ref{pew_he10830_vs_others} we show the pEW measurements 
   of the stars and our models 
   for the \ion{He}{i}~IR line 
   as a function of pEW(\ion{He}{i}~D$_3$) and pEW(H$\alpha$). 
   The models show the \ion{He}{i}~IR line from the state of absorption to 
   being indistinguishable from the surrounding photospheric background, which may indicate a very weak or a filled-in line. 
   In the original model suite, the
   \ion{He}{i}~IR line emission is only found in combination 
   with Pa$\beta$ line emission, which led to their exclusion (Sect.~\ref{Models_flares}). 
   This is in line with observations, where \ion{He}{i}~IR line emission was 
   observed during flares \citep{Schmidt2012ApJ...745...14S, Fuhrmeister2008A&A...487..293F}.  
   
   In the measurements of the \ion{He}{i}~D$_3$ line 
   we found an offset between the observations and the model values. 
   The surrounding continua exhibit a decline from the blue to the red reference bands for the observations, 
   which affects the normalization in the calculations of the pEW values of these lines. 
   Therefore we added an offset of $0.06\,\AA$ to the pEW measurements of the observations. 
   The behavior of the \ion{He}{i}~D$_3$ line is easiest to interpret because it only slightly 
   starts to rise above the continuum with increasing depth 
   of the \ion{He}{i}~IR line for the models as well as for the observations. 
   The series of models created here 
   also shows this trend for all models of series A and for models B1 -- B4, 
   that is, for all the quiescent models of the new series. 
   The flare models B5 and B6 show increasing fill-in of the \ion{He}{i}~IR line, 
   while the line emission of \ion{He}{i}~D$_3$ strengthens further. The same applies to the two most active
   stars of the observed sample. 
   Furthermore, within the inactive range of the models and observations, the \ion{He}{i}~IR 
   sensitivity to the density shift is notably higher than that of the D$_3$ line. 
   This is also confirmed by the respective line evolution shown in Fig.~\ref{series_new_chroms}. 
   This is in line with expectations because the lower level of the D$_3$ line is the 
   upper level of the \ion{He}{i}~IR line. 
   The population of the excited D$_3$ level lags the population of the excited \ion{He}{i}~IR level.

   The H$\alpha$ line behavior in the models shows that no combination of a deep
   \ion{He}{i}~IR line and H$\alpha$ clearly in emission appears to be possible. Nevertheless, the model series A and B and the
   measurements of the observed stars show the same trend: 
   H$\alpha$ behaves similarly to the \ion{He}{i}~IR line: first the line 
   deepens before some fill-in begins that eventually drives the lines into emission. 
   
   We also studied the relationship of pEW(\ion{He}{i}~IR) to pEW(\ion{Na}{i}~D$_2$) 
   and pEW(\ion{Ca}{ii}~IRT). 
   For these lines, 
   the models seem to show no dependence on the
   pEW(\ion{He}{i}~IR). 
   Nevertheless, the observations show that both lines only fill in with increasing activity. 
   For a weak \ion{He}{i}~IR line, measurements of the 
   \ion{Ca}{ii}~IRT or \ion{Na}{i}~D$_2$ line therefore determine if the \ion{He}{i}~IR line
   is intrinsically weak or filled in. 
   
   For all lines, the observed pEW combinations are generally well reproduced 
   by the models. Only the observations of the two active stars \mbox{TYC 3529-1437-1} and 
   \mbox{LP 733-099} show a combination 
   in the pEW(\ion{He}{i}~IR) versus pEW(He~D$_3$) plane that is not covered by the models. 
   Furthermore, the models hardly reproduce the pEW measurements of the stars in the very inactive regime 
   of the pEW(\ion{He}{i}~IR) versus pEW(H$\alpha$) space. 
   For some discussion on these aspects, see Sect.~\ref{sect_prediction}. 
   However, the quiescent models of series A and B mainly 
   follow the observations of the inactive stars.

\subsection{Predicting the \ion{He}{i}~IR line from previous best-fitting models} \label{sect_prediction}

   On the basis of the chromospheric modeling as done in 
   \citet{Hintz2019A&A...623A.136H}, we here determined how well the models reproduce 
   the \ion{He}{i}~IR line. In that study, only 
   the \ion{Na}{i}~D$_2$, H$\alpha$, and the bluest \ion{Ca}{ii}~IRT line were used to identify best-fit models for
   all stars by minimizing a modified $\chi^2$ value ($\chi_m^2$) for these three lines. 
      
   Examples of the inactive stars (with single-model fits) can be found 
   in Fig.~\ref{active_obs_mod_gj70} for GJ~1203 and GJ~625. 
   Here we show the three lines that were used before and 
   the predicted line shape
   of the \ion{He}{i}~IR line, which is reproduced well for GJ~1203 and overpredicted for GJ~625. 
   However, all of our best-fit models show the \ion{He}{i}~IR line in absorption and therefore reproduce
   this characteristic of the observed stellar sample. 
   As an indicator for the goodness of the prediction of the respective models in the range 
   of the \ion{He}{i}~IR line, we calculated reduced $\chi^2$ values in the wavelength interval 
   $10\,833.306\pm0.5\,\AA$ ($\chi^2_{\rm \ion{He}{i}~IR}$). 
   The two different $\chi^2$ quantities, $\chi^2_{\rm \ion{He}{i}~IR}$ and $\chi_m^2$, are not directly comparable.    
   Here we analyze the prediction of the modeled \ion{He}{i}~IR line, 
   therefore we only compare the $\chi^2_{\rm \ion{He}{i}~IR}$ values (see Table~\ref{he_prediction}). 
   From visual inspection, we specified a $\chi^2_{\rm \ion{He}{i}~IR} \le 2$ 
   to flag a model prediction of the \ion{He}{i}~IR line to be adequate for an observation. 
   Approximately for half of the listed stars, 
   the best-fit model prediction is appropriate 
   according to this selection criterion.  
   When the model prediction strongly differs, that is, $\chi^2_{\rm \ion{He}{i}~IR} > 10$, 
   the \ion{He}{i}~IR line absorption is clearly overpredicted, 
   as shown in Fig.~\ref{active_obs_mod_gj70} for GJ~625. 
   
\begin{figure}
   \centering
   \includegraphics[width=0.5\textwidth]{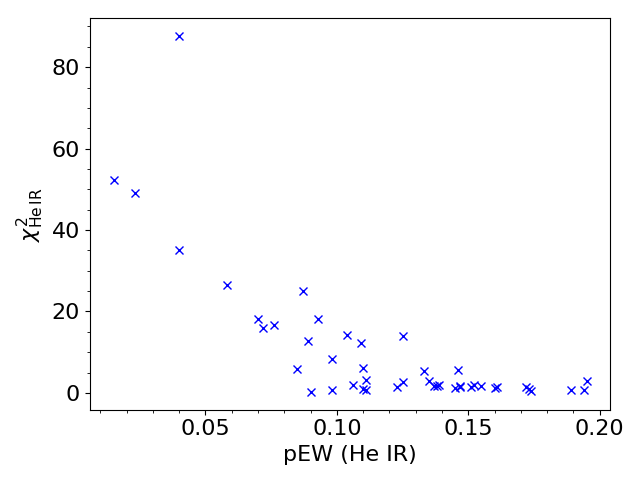}
      \caption{Measurements of the pEW (\ion{He}{i}~IR) plotted against the 
               corresponding $\chi^2_{\rm \ion{He}{i}~IR}$ values for the inactive stars from Table~\ref{he_prediction}. 
              }
         \label{pew_chi_he10830}
   \end{figure}

   Because \citet{Hintz2019A&A...623A.136H} 
   identified only five best-fit models in their 
   the whole sample of inactive stars in our previous study, we now
   determine which of those five models describes the \ion{He}{i}~IR line best and then recompute the
   $\chi_m^2$ values for this model. 
   These results are shown in Table~\ref{he_prediction}. 
   The \ion{He}{i}~IR line is best predicted 
   by models 029 and 047 for all but three stars. 
   These models in many cases describe the \ion{He}{i}~IR line better 
   and show only a marginally worse new $\chi_m^2$ value. A prominent example is GJ~3452. 
   On the other hand, the description of the two most inactive stars, Ross~730 and HD~349726, within the \ion{He}{i}~IR line is
   poor for any of the five best-fit models. This leads us to the general assumption that   
   the \ion{He}{i}~IR line prediction becomes worse 
   when the line is shallow, as is shown in Fig.~\ref{pew_chi_he10830}. 
   This may be caused by a lack of appropriate inactive 
   models in our model suite 
   or may indicate that 
   our ansatz with a 100\% filling factor for the chromospheres fails and even 
   atmospheres of the most inactive stars are partially filled 
   with plage regions. 
   The cases for which the original best fits already yield an adequate \ion{He}{i}~IR line description 
   can by 
   construction lead to an even better description, 
   but it considerably worsens the $\chi_m^2$ 
   in most cases. This shows that more than one 
   individual line needs to be used for chromospheric model fitting, 
   as reported by \citet{Hintz2019A&A...623A.136H}. 
   
\begin{figure*}[h]
   \resizebox{\hsize}{!}
            {\includegraphics[width=0.9\textwidth]{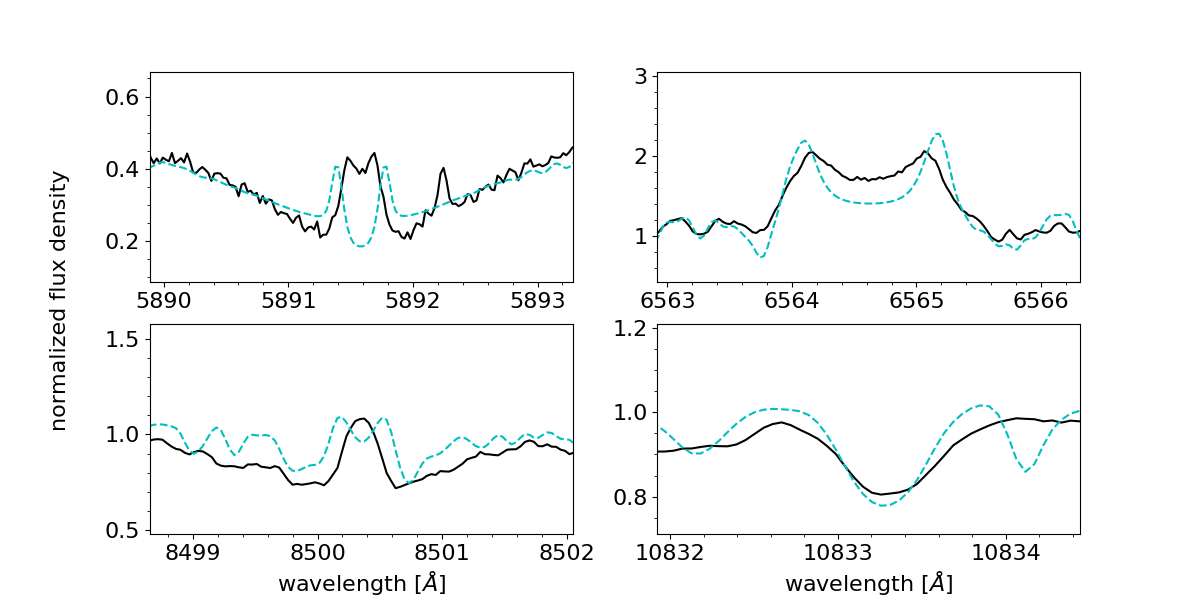}}
      \caption{
               Comparison of the best linear combination fit (dashed cyan line, from \citet{Hintz2019A&A...623A.136H}) 
               to an observed spectrum (solid black) \mbox{\object{TYC~3529-1437-1}} 
               optimized for the lines of \ion{Na}{i}~D$_2$ (\textit{top left panel}), 
               H$\alpha$ (\textit{top right panel}), and the bluest \ion{Ca}{ii}~IRT line  (\textit{bottom left panel}) 
               and extended to the \ion{He}{i}~IR line  (\textit{bottom right panel}). 
               The linear combination consists of 74.9\% of inactive model 079 
               and 25.1\% of active model 149. 
              }
         \label{active_obs_mod_tyc_lin_comb}
   \end{figure*}

   \citet{Hintz2019A&A...623A.136H} found that 
   in the case of the active stars, linear-combination fits of two models (one inactive and one more active model) 
   were necessary to obtain appropriate models. 
   In Fig.~\ref{active_obs_mod_tyc_lin_comb} we illustrate the best linear combination fit for \mbox{TYC~3529-1437-1}. 
   In the combination fits of the active stars, the line in the active component tends to go into emission, 
   but the combination still shows the \ion{He}{i}~IR line in absorption. 
   Table~\ref{table_stars_models_best_comb} lists our information about 
   the \ion{He}{i}~IR line prediction of the combination fits of the active stars. 
   Interestingly, the \ion{He}{i}~IR line can be well reproduced by the best-fit model combinations 
   without including this line in the fitting procedure, 
   as is evident from the $\chi^2_{\rm \ion{He}{i}~IR}$ values.

   \begin{table}[t]
   \caption{Best-fit models for the inactive stars from Table~A.1 of \citet{Hintz2019A&A...623A.136H} 
   and the calculated  reduced $\chi^2$ value (Col. 4) in the \ion{He}{i}~IR line. 
   }            \vspace{-0.7cm}     
   \label{he_prediction}      
   \begin{center}
   \footnotesize
   \begin{tabular}{l c r r c r r}
   \hline\hline       
Stars   & Model   & $\chi_m^2$   & $\chi^2_{\rm \ion{He}{i}~IR}$  & Model$^{\,\,a}$   & $\chi_m^2\,\, ^{a}$   & $\chi^2_{\rm \ion{He}{i}~IR}\,\,^{a}$ \\ 
 \hline 
\object{Wolf 1056}   & 047   & 2.75   & 1.80   & \textcolor{gray}{029}   & \textcolor{gray}{4.73}   & \textcolor{gray}{0.62} \\ 
\object{GJ 47}   & 047   & 2.53   & 1.79   & \textcolor{gray}{029}   & \textcolor{gray}{4.14}   & \textcolor{gray}{0.38} \\ 
\object{BD+70 68}   & 080   & 2.26   & 0.72   & \textcolor{gray}{079}   & \textcolor{gray}{2.74}   & \textcolor{gray}{0.72} \\ 
\object{GJ 70}   & 079   & 2.29   & 1.95   & \textcolor{gray}{029}   & \textcolor{gray}{4.85}   & \textcolor{gray}{0.41} \\ 
\object{G 244-047}   & 042   & 3.45   & 5.73   & 029   & 4.22   & 0.95 \\ 
\object{VX Ari}   & 079   & 3.61   & 6.10   & 047   & 4.11   & 0.73 \\ 
\object{Ross 567}   & 042   & 2.75   & 18.21   & 047   & 3.26   & 0.49 \\ 
\object{GJ 226}   & 047   & 1.80   & 1.67   & \textcolor{gray}{029}   & \textcolor{gray}{4.52}   & \textcolor{gray}{0.48} \\ 
\object{GJ 258}   & 047   & 3.83   & 1.36   & \textcolor{gray}{029}   & \textcolor{gray}{5.27}   & \textcolor{gray}{0.46} \\ 
\object{GJ 1097}   & 042   & 4.04   & 12.79   & 047   & 4.34   & 0.47 \\ 
\object{GJ 3452}   & 042   & 2.37   & 18.09   & 047   & 2.99   & 0.33 \\ 
\object{GJ 357}   & 042   & 2.63   & 35.03   & 047   & 3.25   & 1.85 \\ 
\object{GJ 386}   & 047   & 2.78   & 1.51   & \textcolor{gray}{029}   & \textcolor{gray}{4.00}   & \textcolor{gray}{0.78} \\ 
\object{LP 670-017}   & 080   & 4.51   & 5.92   & 047   & 4.53   & 0.37 \\ 
\object{GJ 399}   & 079   & 2.54   & 1.89   & \textcolor{gray}{029}   & \textcolor{gray}{5.10}   & \textcolor{gray}{0.62} \\ 
\object{Ross 104}   & 079   & 2.59   & 2.58   & 029   & 4.44   & 0.46 \\ 
\object{Ross 905}   & 042   & 2.84   & 12.23   & 047   & 2.94   & 0.75 \\ 
\object{GJ 443}   & 080   & 2.04   & 1.39   & \textcolor{gray}{029}   & \textcolor{gray}{7.48}   & \textcolor{gray}{0.74} \\ 
\object{Ross 690}   & 079   & 2.41   & 5.27   & 047   & 2.81   & 1.31 \\ 
\object{Ross 695}   & 042   & 2.91   & 87.59   & 047   & 3.16   & 5.80 \\ 
\object{Ross 992}   & 079   & 2.85   & 1.67   & \textcolor{gray}{029}   & \textcolor{gray}{5.18}   & \textcolor{gray}{0.42} \\ 
\object{$\theta$ Boo B}   & 047   & 2.55   & 0.78   & \textcolor{gray}{047}   & \textcolor{gray}{2.55}   & \textcolor{gray}{0.78} \\ 
\object{Ross 1047}   & 047   & 3.16   & 0.94   & \textcolor{gray}{029}   & \textcolor{gray}{4.41}   & \textcolor{gray}{0.72} \\ 
\object{LP 743-031}   & 080   & 3.38   & 14.27   & 047   & 4.69   & 0.97 \\ 
\object{G 137-084}   & 080   & 2.66   & 1.96   & \textcolor{gray}{029}   & \textcolor{gray}{4.97}   & \textcolor{gray}{0.36} \\ 
\object{EW Dra}   & 080   & 2.31   & 0.46   & \textcolor{gray}{080}   & \textcolor{gray}{2.31}   & \textcolor{gray}{0.46} \\ 
\object{GJ 625}   & 042   & 2.41   & 25.02   & 047   & 2.43   & 0.63 \\ 
\object{GJ 1203}   & 047   & 2.50   & 0.24   & \textcolor{gray}{047}   & \textcolor{gray}{2.50}   & \textcolor{gray}{0.24} \\ 
\object{LP 446-006}   & 047   & 2.82   & 1.11   & \textcolor{gray}{029}   & \textcolor{gray}{3.72}   & \textcolor{gray}{1.03} \\ 
\object{Ross 863}   & 079   & 3.11   & 3.08   & 029   & 4.02   & 0.60 \\ 
\object{GJ 2128}   & 042   & 2.45   & 16.69   & 047   & 2.58   & 0.41 \\ 
\object{GJ 671}   & 042   & 3.30   & 15.92   & 047   & 3.55   & 0.31 \\ 
\object{G 204-039}   & 080   & 2.79   & 1.34   & \textcolor{gray}{029}   & \textcolor{gray}{5.82}   & \textcolor{gray}{0.38} \\ 
\object{Ross 145}   & 042   & 3.28   & 26.56   & 047   & 4.19   & 0.71 \\ 
\object{G 155-042}   & 042   & 3.54   & 8.40   & 047   & 3.56   & 0.76 \\ 
\object{Ross 730}   & 029   & 2.82   & 52.27   & 047   & 6.02   & 15.92 \\ 
\object{HD 349726}   & 029   & 2.83   & 49.21   & 047   & 5.61   & 14.22 \\ 
\object{GJ 793}   & 080   & 3.01   & 1.17   & \textcolor{gray}{029}   & \textcolor{gray}{7.02}   & \textcolor{gray}{0.37} \\ 
\object{Wolf 896}   & 047   & 2.59   & 2.83   & 079   & 2.66   & 0.73 \\ 
\object{Wolf 906}   & 079   & 2.43   & 0.92   & \textcolor{gray}{029}   & \textcolor{gray}{5.29}   & \textcolor{gray}{0.65} \\ 
LSPM J2116 &    &    &    &    &    &  \\ 
\, +0234   & 079   & 3.08   & 1.54   & \textcolor{gray}{029}   & \textcolor{gray}{5.10}   & \textcolor{gray}{0.57} \\ 
\object{BD-05 5715}   & 080   & 2.90   & 0.69   & \textcolor{gray}{080}   & \textcolor{gray}{2.90}   & \textcolor{gray}{0.69} \\ 
\object{Wolf 1014}   & 042   & 3.32   & 13.91   & 047   & 4.21   & 0.94 \\ 
\object{G 273-093}   & 047   & 1.97   & 0.74   & \textcolor{gray}{047}   & \textcolor{gray}{1.97}   & \textcolor{gray}{0.74} \\ 
\object{Wolf 1051}   & 080   & 2.31   & 2.94   & 029   & 6.39   & 0.68 \\ 
   \hline   
   \end{tabular}
   \end{center}
   \footnotesize{$^{a}$ Columns 5--7 list the results of the model with the best \ion{He}{i}~IR line prediction 
   within the subsample of the five best-fitting models for the inactive stars. 
   For stars whose original model already resulted in an adequate 
   description of the \ion{He}{i}~IR line, we mark the \ion{He}{i}~IR line best-fit model in gray.} 
   \end{table}

   \begin{table}
   \caption{Best-fit models of the active stars in a linear-combination fit with filling factors (FF) and 
   the calculated $\chi^2$ value (Col. 5) in the \ion{He}{i}~IR line.}             
   \label{table_stars_models_best_comb}      
   \centering
   \footnotesize
   \begin{tabular}{l c c c c c c}
   \hline\hline       
Stars   & Inactive  & FF & Active & FF & $\chi^2_m$ & $\chi^2_{\rm \ion{He}{i}~IR}$ \\ 
        & model &  & model &  &  &  \\ 
 \hline 
\object{G 234-057}   & 080   & 0.90   & 139   & 0.10   & 3.58   & 1.54 \\ 
\object{GJ 360}   & 080   & 0.82   & 132   & 0.17   & 5.13   & 0.70 \\ 
\object{LP 733-099}   & 079   & 0.68   & 149   & 0.32   & 16.13   & 0.48 \\ 
TYC 3529-&    &    &    &    &    &  \\ 
\, 1437-1   & 079   & 0.75   & 149   & 0.25   & 14.25   & 0.50 \\ 
   \hline   
   \end{tabular} 
   \end{table}

\section{Summary and conclusions}
  
   We presented a theoretical study of the \ion{He}{i}~IR line in M2--3\,V stars
   with PHOENIX. Our study was based on a set of chromospheric 
   PHOENIX models 
   computed in our previous study, \citet{Hintz2019A&A...623A.136H}, and new systematic
   series of chromospheric models. 
   Following the approach 
   of previous M-dwarf star studies, 
   all our models (previous and new models) were created 
   by parameterizing the 
   semiempirical solar VAL~C temperature structure. 
   The model spectra were then used 
   to fit observed CARMENES spectra of M2--3\,V stars in 
   the \ion{Na}{i}~D$_2$, H$\alpha$, and the bluest \ion{Ca}{ii}~IRT lines. 
   A statistical analysis reveals the best fits. 
   Here, we extended the study to investigate the behavior of the 
   \ion{He}{i}~IR line. 

   Our new model series were computed by shifting
   an inactive model structure in column mass density. In these series,
   denser chromospheres correspond to higher activity levels.
   From low to high activity levels, our models predict that the
   \ion{He}{i}~IR line first goes into absorption and then reaches a maximum
   depth, after which fill-in sets in. 
   Our models reproduce the point of maximum \ion{He}{i}~IR line absorption at $\sim 300$\,m\AA. 
   The H$\alpha$ line shows the same qualitative behavior, but reaches the turning
   point at lower activity levels. 
   In contrast, the \ion{Na}{i}~D$_2$ and the bluest \ion{Ca}{ii}~IRT lines
   only show fill-in with increasing activity levels. 
   Furthermore, the \ion{He}{i}~D$_3$ line never shows absorption in our models, but
   directly goes into emission with increasing activity levels. 
   
   By investigating the \ion{He}{i}~IR line as a function of the EUV radiation field arising from the models themselves, 
   we find that the most inactive models are highly sensitive to an increase in EUV radiation, 
   which produces a strong rise in the pEW(\ion{He}{i}~IR). 
   With further increases of the EUV radiation, the line absorption tends to saturate. 
   The detailed response of the radiation field and the \ion{He}{i}~IR line to 
   density shifts of the atmospheric structure depends on the 
   configuration of the chromosphere. 
   
   Suppressing EUV emission and thus the PR mechanism altogether, we showed that
   collisional excitation gains in importance for the formation of the \ion{He}{i}~IR
   line as activity levels increase beyond
   the point of maximum \ion{He}{i}~IR line absorption.
   Our investigation of chromosphere models showed  
   that the PR mechanism and collisional excitation both contribute to the \ion{He}{i}~IR
   formation, with the PR mechanism dominating the low-activity regime.
   This is in line with former results 
   on the \ion{He}{i}~IR line formation in the Sun. 
   
   The whole stellar comparison sample 
   shows the \ion{He}{i}~IR line in absorption. 
   We showed that the best-fit models for inactive stars selected by \citet{Hintz2019A&A...623A.136H}
   based on optical activity indicators often also predict the \ion{He}{i}~IR line
   quite well. In other cases, an appropriate inactive model can be found that fits
   both the optical and \ion{He}{i}~IR lines well. 
   For active stars, a linear combination of an active and inactive component
   is required to obtain a good approximation, as reported in \citet{Hintz2019A&A...623A.136H}.
   
   Our current study demonstrates that one-dimensional PHOENIX model atmospheres with a parameterized temperature
   structure reproduce the observed behavior of the \ion{He}{i}~IR line in M2--3\,V stars.
   Its strong response to EUV irradiation in the low-activity regime makes the \ion{He}{i}~IR line
   a promising proxy of this otherwise inaccessible wavelength range there.
   The comparably weak response of the line to changes in the EUV radiation field that start already at
   moderate activity levels is favorable for planetary transmission spectroscopy of active stars, where
   constant reference spectra are crucial.

\begin{acknowledgements}
   CARMENES is an instrument for the Centro Astron\'omico Hispano-Alem\'an 
   de
   Calar Alto (CAHA, Almer\'{\i}a, Spain).
   CARMENES is funded by the German Max-Planck-Gesellschaft (MPG), 
   the Spanish Consejo Superior de Investigaciones Cient\'{\i}ficas (CSIC),
   the European Union through FEDER/ERF FICTS-2011-02 funds,
   and the members of the CARMENES Consortium
   (Max-Planck-Institut f\"ur Astronomie, 
   Instituto de Astrof\'{\i}sica de Andaluc\'{\i}a, 
   Landessternwarte K\"onigstuhl, 
   Institut de Ci\`encies de l'Espai, 
   Institut f\"ur Astrophysik G\"ottingen, 
   Universidad Complutense de Madrid, 
   Th\"uringer Landessternwarte Tautenburg, 
   Instituto de Astrof\'{\i}sica de Canarias, 
   Hamburger Sternwarte, 
   Centro de Astrobiolog\'{\i}a and
   Centro Astron\'omico Hispano-Alem\'an), 
   with additional contributions by the Spanish Ministry of 
   Science  [through  projects  AYA2016-79425-C3-1/2/3-P, 
   ESP2016-80435-C2-1-R, AYA2015-69350-C3-2-P, and AYA2018-84089],
   the German Science Foundation through the Major Research Instrumentation
   Programme and DFG Research Unit FOR2544 ``Blue Planets around Red 
   Stars'',
   the Klaus Tschira Stiftung, 
   the states of Baden-W\"urttemberg and Niedersachsen, 
   and by the Junta de Andaluc\'{\i}a. 
   D.H. acknowledges funding by the DLR under DLR 50 OR1701. B.F. 
   acknowledges funding by the DFG under Cz \mbox{222/1-1} and Schm 
   1032/69-1. 
   S.C. acknowledges support through DFG projects SCH 1382/2-1 and SCHM 1032/66-1.
   We thank the anonymous referee for the comprehensive review.
\end{acknowledgements}

\bibliographystyle{aa} 
\bibliography{references}

\begin{appendix} 

\section{Reviewed former model set}
Here, we list the parameters of the models from \citet{Hintz2019A&A...623A.136H} that we review in this work 
according to our selection criteria in Sect.~\ref{Models_flares}. 
The non-listed models either show the line of Pa$\beta$ in obvious emission 
or their onset of the transition region falls below a temperature of $T_{\rm top} = 6000\,$K. 

\begin{table}[h!] 
\label{table_model_grid}  
\caption{Reviewed former model set and parameters 
from Table~C.1 in \citet[][]{Hintz2019A&A...623A.136H} for the models 
that fulfill our selection criteria as described in Sect.~\ref{Models_flares}.} 
\centering
\begin{tabular}{c c c c c c c}   
\hline\hline     
Model & $m_\mathrm{min}$ & $m_\mathrm{mid}$ & $T_\mathrm{mid}$ & $m_\mathrm{top}$ & $T_\mathrm{top}$ & $\nabla_{\mathrm{TR}}$ \\ 
      & $[\mathrm{dex}]$ & $[\mathrm{dex}]$ & $[\mathrm{K}]$   & $[\mathrm{dex}]$ & $[\mathrm{K}]$   & $[\mathrm{dex}]$ \\ 
\hline

001   & -4.0   & -4.3   & 5500   & -6.0   & 6000   & 7.5  \\ 
005   & -3.5   & -3.8   & 5500   & -5.5   & 6000   & 7.5  \\ 
009   & -3.2   & -3.6   & 5500   & -5.1   & 6000   & 7.5  \\ 
014   & -3.1   & -3.3   & 5500   & -5.0   & 6000   & 8.5  \\ 
016   & -3.0   & -3.6   & 5500   & -5.0   & 6000   & 7.5  \\ 
020   & -3.0   & -3.3   & 5500   & -5.0   & 6000   & 7.5  \\ 
021   & -3.0   & -3.3   & 5500   & -5.0   & 6000   & 7.5  \\ 
023   & -2.8   & -3.6   & 5500   & -5.0   & 6000   & 7.5  \\ 
025   & -2.8   & -3.3   & 5500   & -5.0   & 6000   & 7.5  \\ 
028   & -2.6   & -3.2   & 4500   & -4.5   & 6000   & 8.5  \\ 
029   & -2.6   & -3.2   & 4500   & -4.5   & 7000   & 8.5  \\ 
031   & -2.6   & -3.0   & 4500   & -4.5   & 6000   & 8.5  \\ 
032   & -2.6   & -3.0   & 4500   & -4.5   & 7000   & 8.5  \\ 
037   & -2.6   & -2.8   & 4500   & -4.5   & 6000   & 8.5  \\ 
038   & -2.6   & -2.8   & 4500   & -4.5   & 7000   & 8.5  \\ 
040   & -2.5   & -3.6   & 5500   & -5.0   & 6000   & 7.5  \\ 
041   & -2.5   & -3.6   & 5500   & -5.0   & 6200   & 7.5  \\ 
042   & -2.5   & -3.6   & 5500   & -5.0   & 6500   & 7.5  \\ 
044   & -2.5   & -3.3   & 5500   & -5.0   & 6000   & 7.5  \\ 
046   & -2.5   & -2.8   & 5500   & -4.5   & 6000   & 7.5  \\ 
047   & -2.5   & -2.7   & 6500   & -5.0   & 7000   & 9.2  \\ 
049   & -2.1   & -2.6   & 6500   & -4.5   & 7000   & 7.5  \\ 
050   & -2.1   & -2.6   & 6500   & -4.0   & 7000   & 9.2  \\ 
055   & -2.1   & -2.3   & 5000   & -4.0   & 6000   & 9.5  \\ 
056   & -2.1   & -2.3   & 5500   & -5.0   & 6000   & 7.5  \\ 
057   & -2.1   & -2.3   & 5500   & -4.5   & 6000   & 7.5  \\ 
059   & -2.1   & -2.3   & 5500   & -4.0   & 6000   & 9.5  \\ 
060   & -2.1   & -2.3   & 6500   & -5.0   & 7000   & 9.2  \\ 
061   & -2.1   & -2.3   & 6500   & -4.5   & 7000   & 9.2  \\ 
062   & -2.1   & -2.3   & 6500   & -4.2   & 7000   & 8.2  \\ 
063   & -2.1   & -2.3   & 6500   & -4.0   & 7000   & 9.0  \\ 
064   & -2.1   & -2.3   & 6500   & -4.0   & 7000   & 9.2  \\ 
065   & -2.0   & -2.5   & 6500   & -5.0   & 8000   & 9.2  \\ 
066   & -2.0   & -2.5   & 6500   & -4.5   & 8000   & 9.2  \\ 
068   & -2.0   & -2.3   & 6500   & -5.0   & 7000   & 9.2  \\ 
069   & -2.0   & -2.3   & 6500   & -4.5   & 7000   & 9.2  \\ 
070   & -2.0   & -2.3   & 6500   & -4.0   & 7000   & 9.2  \\ 
071   & -2.0   & -2.2   & 6500   & -5.0   & 7000   & 9.2  \\ 
072   & -2.0   & -2.2   & 6500   & -4.5   & 7000   & 9.2  \\ 
073   & -1.9   & -2.3   & 6500   & -4.0   & 7000   & 9.2  \\ 
078   & -1.8   & -2.3   & 6500   & -4.0   & 7000   & 9.0  \\ 
079   & -1.5   & -2.5   & 5000   & -4.0   & 7500   & 8.5  \\ 
080   & -1.5   & -2.5   & 5500   & -4.0   & 7500   & 8.5  \\ 
081   & -1.5   & -2.5   & 6000   & -4.1   & 7500   & 8.5  \\ 
083   & -1.5   & -2.5   & 6000   & -4.0   & 7500   & 8.5  \\ 
095   & -1.5   & -2.3   & 6500   & -4.5   & 7000   & 9.0  \\ 
096   & -1.5   & -2.3   & 6500   & -4.5   & 8000   & 9.2  \\ 
097   & -1.5   & -2.3   & 6500   & -4.0   & 7000   & 9.0  \\ 
107   & -1.0   & -3.5   & 4000   & -5.0   & 8000   & 9.5  \\ 
108   & -1.0   & -3.5   & 4000   & -4.5   & 8000   & 9.5  \\ 
109   & -1.0   & -3.0   & 4000   & -5.0   & 8000   & 9.5  \\ 
110   & -1.0   & -3.0   & 4000   & -4.5   & 8000   & 9.5  \\ 
111   & -1.0   & -3.0   & 4000   & -4.0   & 8000   & 9.5  \\ 
124   & -1.0   & -2.5   & 4000   & -4.0   & 8000   & 9.5  \\ 
131   & -1.0   & -2.5   & 5200   & -3.7   & 8000   & 9.5  \\ 
152   & -1.0   & -2.0   & 4000   & -4.0   & 8000   & 9.5  \\ 
155   & -1.0   & -1.5   & 4000   & -4.0   & 8000   & 9.5  \\ 
159   & -0.3   & -3.0   & 4000   & -4.5   & 8000   & 9.5  \\ 
\hline
\end{tabular}
\end{table}

\end{appendix}

\end{document}